\documentclass[%
 reprint,%
 amssymb, amsmath,%
 aip,cha,%
]{revtex4-1}

\usepackage{bm}%
\usepackage{graphicx}
\usepackage[colorlinks=true,linkcolor=blue]{hyperref}%
\expandafter\ifx\csname package@font\endcsname\relax\else
 \expandafter\expandafter
 \expandafter\usepackage
 \expandafter\expandafter
 \expandafter{\csname package@font\endcsname}%
\fi
\newcommand{\bra}[1]{\ensuremath{\left\langle{#1}\right\vert}}

\newcommand{\ket}[1]{\ensuremath{\left|{#1}\right\rangle}}

\newcommand{\abs}[1]{\left|#1\right|}

\begin{document}

\title{Signatures of spatially correlated noise and non-secular effects\\in two-dimensional electronic spectroscopy}%

\author{James Lim}\thanks{These authors contributed equally to this work.}%
\affiliation{Institut f\"ur Theoretische Physik, Albert-Einstein-Allee 11, Universit\"at Ulm, D-89069 Ulm, Germany}%

\author{David J. Ing}\thanks{These authors contributed equally to this work.}%
\affiliation{Chemical and Quantum Physics, School of Applied Sciences, RMIT University, Melbourne, Victoria 3001, Australia}%

\author{Joachim Rosskopf}%
\affiliation{Institut f\"ur Theoretische Physik, Albert-Einstein-Allee 11, Universit\"at Ulm, D-89069 Ulm, Germany}%

\author{Jan Jeske}%
\affiliation{Chemical and Quantum Physics, School of Applied Sciences, RMIT University, Melbourne, Victoria 3001, Australia}%

\author{Jared H. Cole}%
\affiliation{Chemical and Quantum Physics, School of Applied Sciences, RMIT University, Melbourne, Victoria 3001, Australia}%

\author{Susana F. Huelga}%
\email{susana.huelga@uni-ulm.de}
\affiliation{Institut f\"ur Theoretische Physik, Albert-Einstein-Allee 11, Universit\"at Ulm, D-89069 Ulm, Germany}%

\author{Martin B. Plenio}%
\email{martin.plenio@uni-ulm.de}
\affiliation{Institut f\"ur Theoretische Physik, Albert-Einstein-Allee 11, Universit\"at Ulm, D-89069 Ulm, Germany}%

\begin{abstract}
We investigate how correlated fluctuations affect oscillatory features in rephasing and non-rephasing two-dimensional (2D) electronic spectra of a model dimer system. Based on a beating map analysis, we show that non-secular environmental couplings induced by uncorrelated fluctuations lead to oscillations centered at both cross- and diagonal-peaks in rephasing spectra as well as in non-rephasing spectra. Using an analytical approach, we provide a quantitative description of the non-secular effects in terms of the Feynman diagrams and show that the environment-induced mixing of different inter-excitonic coherences leads to oscillations in the rephasing diagonal-peaks and non-rephasing cross-peaks. We demonstrate that as correlations in the noise increase, the lifetime of oscillatory 2D signals is enhanced at rephasing cross-peaks and non-rephasing diagonal-peaks, while the other non-secular oscillatory signals are suppressed. We discuss that the asymmetry of 2D lineshapes in the beating map provides information on the degree of correlations in environmental fluctuations. Finally we investigate how the oscillatory features in 2D spectra are affected by inhomogeneous broadening.
\end{abstract}

\maketitle


\section{Introduction}

In the first step of the light-harvesting process~\cite{Amerogen2000,Blankenship2002}, the neutral electronic excitations (excitons) created by light absorption are transferred through the molecular system until free charge carriers are generated by exciton dissociation~\cite{Amerogen2000,Blankenship2002}. The interaction between electronic and vibrational degrees of freedom governs the exciton transfer dynamics, such as coherent and incoherent features in energy transport~\cite{IshizakiPCCP2010,HuelgaCP2013,ChenuARPC2015}.

Two-dimensional electronic spectroscopy (2DES) has been employed to study the exciton transfer dynamics in the light-harvesting systems on a sub-picosecond timescale~\cite{JonasARPC2003}. For various natural~\cite{Engel2007,Collini2010,CaramJCP2012,FidlerJPCA2012,RomeroNP2014,Fuller2014} and artificial~\cite{MilotaJPCA2013,HayesScience2013,HalpinNC2014,SongNC2014,LimNC2015,CassetteNC2015,BolzonelloJPCL2016,SioArxiv2016} systems, oscillatory signals observed in 2D experiments have been interpreted as a signature of quantum coherences within the molecular system, coherence that is generated by laser pulses but sustained by the intrinsic dynamics. These coherences can originate in principle from both electronic and vibrational degrees of freedom. The electronic states of the light-harvesting systems are coupled to their vibrational environments with a moderate coupling strength, such that electronic coherences are not completely washed out by the environment-induced noise. The intra-pigment vibrations of the light-harvesting systems exhibit underdamped vibrational motions on a picosecond timescale, which can leads to vibrational coherences in the electronic ground state manifold~\cite{ButkusCPL2012,TiwariPNAS2013}. A vibronic (electronic-vibrational) coupling between electronic states and underdamped vibrational motions leads to a mixing of electronic and vibrational degrees of freedom, inducing vibronic coherences in the electronic excited state manifold~\cite{PriorPRL2010,ChinNJP2010,ChristenssonJPCB2012,ChinNP2013,TiwariPNAS2013,PlenioJCP2013,ChenuSR2013,ButkusJCP2014,BasinskaitePR2014}. A model dimer system has been widely employed to investigate how electronic-vibrational interactions are reflected in spectroscopy~\cite{PlenioJCP2013,ChenuSR2013,ButkusJCP2014,BasinskaitePR2014,YangJMS2006,PolyutovCP2012,SchroterPR2015}.

For various light-harvesting systems, 2DES has demonstrated the presence of long-lived quantum coherences, which are sustained beyond the lifetime of optical coherences between electronic ground and excited states \cite{Engel2007,Collini2010,CaramJCP2012,FidlerJPCA2012,RomeroNP2014,Fuller2014,MilotaJPCA2013,HayesScience2013,HalpinNC2014,SongNC2014,LimNC2015}. To identify the microscopic origin of the long-lived oscillatory signals observed in 2D experiments, several hypotheses have been formulated to explain how quantum coherences are sustained under a noisy environment at ambient conditions. Coherent vibronic coupling has been shown to induce long-lived vibronic coherences when the vibrational frequency is resonant with the energy-level difference between exciton states~\cite{PriorPRL2010,ChinNJP2010,ChristenssonJPCB2012,ChinNP2013,TiwariPNAS2013,PlenioJCP2013,ChenuSR2013} and the vibronic coupling strength is moderate in magnitude~\cite{PlenioJCP2013,LimNC2015}. In this case, oscillatory 2D signals originate from the combination of excited-state vibronic coherence and ground-state vibrational coherence, as they originate from the same mechanism, namely the vibronic Hamiltonian. Recent 2D experiments on J-aggregates of cyanine dyes \cite{LimNC2015} confirmed the validity of this theoretical approach.  Other experimental results on a synthetic dimer \cite{HalpinNC2014,DuanNJP2015} pointed towards the need of a threshold in the vibronic coupling strength to allow vibronic mixing to be relevant under environmental effects.

Correlated fluctuations in the transition energies of neighboring pigments have been suggested as an alternative mechanism where purely electronic coherences in the excited state manifold induce long-lived oscillatory 2D signals~\cite{LeeScience2007,IshizakiPCCP2010,JeskeJCP2015}. In the correlated fluctuation model, where underdamped vibrational motions are not considered, uncorrelated noise between electronic ground and excited states leads to the finite homogeneous broadening of 2D spectra, while highly correlated noise between different excited states induces long-lived inter-excitonic coherences. This purely electronic coherence model is similar in spirit to the decoherence-free subspaces in quantum information science~\cite{LidarPRL1998,JeskePRA2013_2}. Correlated fluctuations have been shown to enhance the lifetime of coherent features in electronic motions, such as population dynamics~\cite{NalbachNJP2010,IshizakiNJP2010,ChenJCP2010,McCutcheonPRB2011,LimNJP2014,JeskeJCP2015}. Correlated fluctuations have also been considered in the simulations of 2D electronic spectroscopy~\cite{AbramaviciusJCP2011,SeibtJCP2014}. Using the secular Redfield theory, it was found that uncorrelated and correlated fluctuation models give very similar absorption spectra and rephasing diagonal-peaks, but show large differences in rephasing cross-peak regimes~\cite{AbramaviciusJCP2011}. The enhancement of the lifetime of 2D oscillations was observed for rephasing diagonal- and cross-peaks, but the diagonal-peak oscillations were attributed to the overlap of non-oscillatory diagonal-peaks with nearby oscillating cross-peaks~\cite{AbramaviciusJCP2011}. It was found that correlations in the noise can alter the relative phase between diagonal- and cross-peak oscillations~\cite{SeibtJCP2014}. Correlations in inhomogeneous broadening have also been considered in the context of 2DES where correlated and anti-correlated static disorder lead to cross-peaks strongly elongated along diagonal and anti-diagonal directions, respectively~\cite{RancovaJCP2015}. A similar feature was observed in the simulations of 2D infrared (IR) spectroscopy in the slow bath limit, while different correlation models lead to similar Lorentzian 2D lineshapes in the fast bath limit~\cite{VenkatramaniJCP2002}. However, it is unclear what the microscopic origin of such correlations is.  Quantum mechanics/molecular mechanics (QM/MM) simulations of photosynthetic systems, such as the Fenna-Matthews-Olson (FMO) complex \cite{OlbrichJPCB2011,ShimBJ2012} and phycoerythrin 545 (PE545) complex \cite{VianiJPCL2013}, for instance, have shown no evidence of spatially correlated fluctuations.  This is contrary to the 2D experiments on the FMO complex \cite{CaramJCP2012,FidlerJPCA2012}, colloidal semiconductor nanoplatelets \cite{CassetteNC2015} and J-aggregates of porphyrins~\cite{BolzonelloJPCL2016}, where the presence of correlated fluctuations was suggested.  The discrepancy between theory and experiment shows the need for further investigations of how the degree of correlations in the noise is reflected in experimental observables, so that the presence of correlated fluctuations can be verified or ruled out based upon experimental results.

In this work, we employ the Bloch-Redfield equation~\cite{JeskePRA2013,JeskeJCP2015} where the degree of correlations in the noise is parameterized by a continuous variable to investigate how the correlations affect the oscillatory features in rephasing and non-rephasing 2D spectra of a model dimer system. We do not include underdamped vibrational modes in the model to focus on the influence of spatial noise correlations on optical responses without vibronic effects. The employed Bloch-Redfield equation includes non-secular environmental couplings between exciton populations and inter-excitonic coherences, and also the non-secular interaction between different inter-excitonic coherences. These non-secular terms are disregarded in the secular approximation, where the dynamics of exciton populations are decoupled from those of inter-excitonic coherences.  The secular terms describe the relaxation between exciton populations and the decay of inter-excitonic coherences, independently.

We show that in the absence of noise correlations, non-secular environmental couplings induce oscillatory 2D signals centered at both cross- and diagonal-peaks in rephasing spectra as well as in non-rephasing spectra. In Ref.~\onlinecite{AbramaviciusJCP2010}, it was found that rephasing diagonal-peaks show notable oscillatory features when non-secular couplings are taken into account in simulations. The origin of the oscillations was attributed to the non-secular interaction between exciton populations and coherences~\cite{AbramaviciusJCP2010,PanitchayangkoonPNAS2011}. It was also shown that the diagonal-peak oscillations are suppressed when the secular approximation is employed, even though both diagonal- and cross-peaks showed similar amplitudes of oscillations~\cite{AbramaviciusJCP2010}. In this work, to clarify the contribution of non-secular effects to rephasing diagonal- and non-rephrasing cross-peak oscillations (non-secular oscillations) in the presence of spectral overlap of diagonal- and cross-peaks, we provide a {\it quantitative} beating map analysis in terms of the eigenstates of the Liouville space operator with associated Feynman diagrams. The beating map analysis helps to clarify whether the oscillation of a given 2D peak originates from itself or merely from the overlap with nearby oscillating peaks. For a homodimer, we show that the non-secular oscillations can be induced by the non-secular coupling between different inter-excitonic coherences, even if their dynamics are decoupled from those of exciton populations. For a heterodimer, where the dynamics of exciton populations and coherences are all coupled to one another, we show that the non-secular oscillations are dominated by the mixing of different inter-excitonic coherences, rather than population-coherence mixing suggested in Refs.~\onlinecite{AbramaviciusJCP2010,PanitchayangkoonPNAS2011}. In addition, we show that the uncorrelated noise can induce asymmetric lineshapes of homogeneously broadened 2D peaks in the beating map, elongated along excitation or detection axis, when the broadening is dominated by relaxation, rather than pure dephasing noise. For the FMO complex, small pure dephasing rates have been identified as a condition for the oscillations of rephasing cross-peaks \cite{KreisbeckJPCL2012}. Finally we show that as the degree of correlations in the noise increases, the lineshape of absorption and 2D electronic spectra, including both diagonal- and cross-peaks, is significantly changed, as the homogeneous broadening starts to be dominated by pure dephasing, rather than relaxation. We show that the noise correlations induce long-lasting oscillations in the rephasing cross-peaks and non-rephasing diagonal-peaks with suppressed non-secular oscillations, leading to symmetric 2D lineshapes in the beating map. We discuss that the asymmetry of 2D lineshapes can provide information on the degree of correlated fluctuations in J-aggregates~\cite{LimNC2015} and colloidal semiconductor nanoplatelets~\cite{CassetteNC2015}, for which asymmetric lineshapes elongated along the excitation axis were observed in 2D experiments.


\section{The Model}

\begin{figure}[t]
	\includegraphics{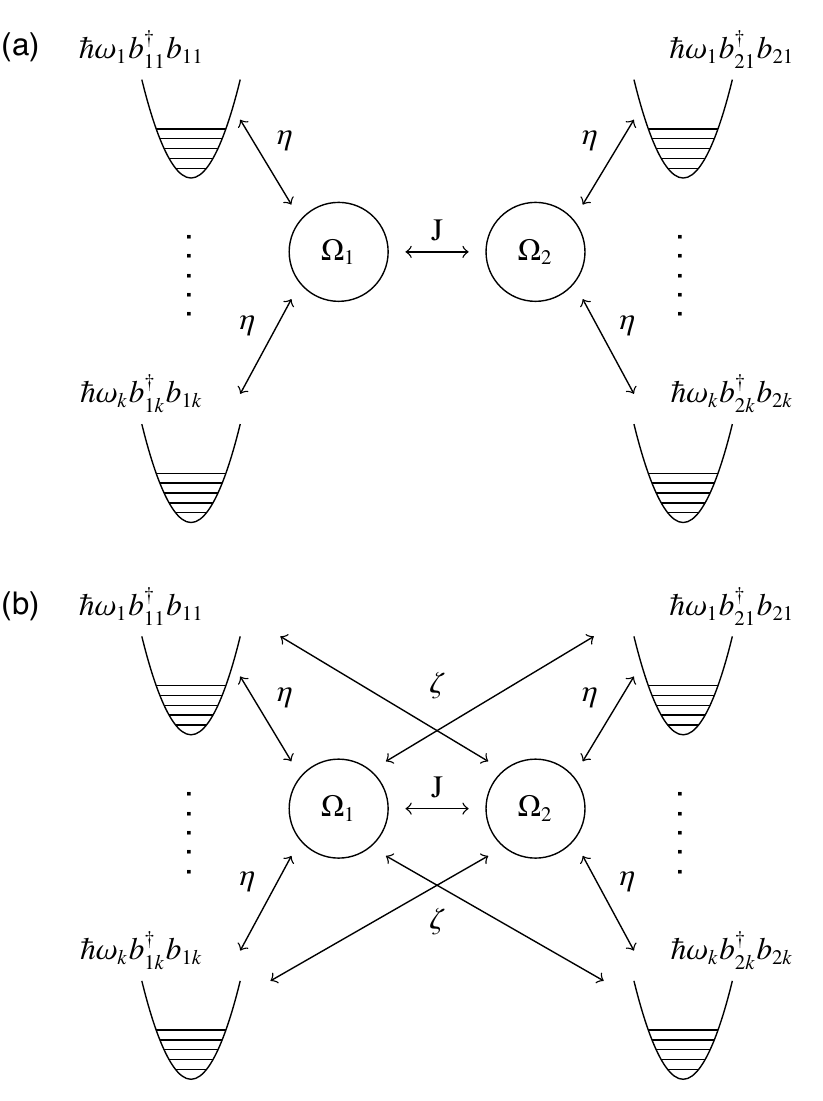}
	\caption{A schematic representation of (a) local phonon baths and (b) a shared phonon bath. In (a), a phonon mode $b_{1k}$ is locally coupled to site 1. In (b), a phonon mode $b_{1k}$ is coupled to both sites 1 and 2 with the relative electron-phonon coupling strengths quantified by $\eta$ and $\zeta$, respectively.}
	\label{figureindex1} 
\end{figure}

To investigate how correlations in the noise are reflected in two-dimensional electronic spectroscopy, we consider a dimer system consisting of two sites coherently coupled by an electronic coupling $J$. The Hamiltonian of the dimer system is modeled by
\begin{equation}
	H_e=\hbar\Omega_1\sigma_{1}^{+}\sigma_{1}^{-}+\hbar\Omega_2\sigma_{2}^{+}\sigma^-_{2}+\hbar J(\sigma_{1}^{+}\sigma^-_{2}+\sigma_{1}^{-}\sigma^+_{2}),
\end{equation}
with $\Omega_k$ denoting the energy level of the site $k$, $\sigma^-_{1}=\ket{g_1}\bra{e_1}\otimes\openone_2$ and $\sigma^-_{2}=\openone_1\otimes\ket{g_2}\bra{e_2}$ represent the annihilation operators of the electronic excitation at sites 1 and 2, respectively, with $\ket{g_k}$ and $\ket{e_k}$ denoting the ground and excited states of the site $k$, respectively, with $\openone_k=\ket{g_k}\bra{g_k}+\ket{e_k}\bra{e_k}$. The electronic eigenstates of $H_e$ are expressed as
 \begin{align}
\ket{g}&=\ket{g_1,g_2},\\
\ket{\epsilon_1}&=-\sin(\theta)\ket{e_1,g_2}+\cos(\theta)\ket{g_1,e_2},\label{eq:evec1}\\
\ket{\epsilon_2}&=\cos(\theta)\ket{e_1,g_2}+\sin(\theta)\ket{g_1,e_2},\label{eq:evec2}\\
\ket{f}&=\ket{e_1,e_2},
 \end{align}
 with $\theta=\frac{1}{2}\tan^{-1}(2J/(\Omega_1-\Omega_2))$ for $\Omega_1\ge\Omega_2$. The associated eigenvalues are given by
 \begin{align}
\epsilon_g&=0,\\
\epsilon_1&=\frac{1}{2}\left(\Omega_1+\Omega_2-\sqrt{(\Omega_1-\Omega_2)^{2}+4J^{2}}\right),\label{eq:e1}\\
\epsilon_2&=\frac{1}{2}\left(\Omega_1+\Omega_2+\sqrt{(\Omega_1-\Omega_2)^{2}+4J^{2}}\right),\label{eq:e2}\\
\epsilon_f&=\Omega_1+\Omega_2,
 \end{align}
 where $\epsilon_1<\epsilon_2$ and the bi-exciton binding energy is not considered for the sake of simplicity, leading to $\epsilon_f=\epsilon_1+\epsilon_2$. Here $\ket{g}$ represents a common ground state, $\ket{\epsilon_1}$ and $\ket{\epsilon_2}$ are low and high energy single exciton states, respectively, and $\ket{f}$ is a bi-exciton state.

The phonon environment coupled to the electronic states is modeled by independent harmonic oscillators
\begin{equation}
	H_p=\sum_{k}(\hbar\omega_{k}b_{1k}^{\dagger}b_{1k}+\hbar\omega_{k}b_{2k}^{\dagger}b_{2k}),
\end{equation}
where the dephasing interaction between electronic states and phonon environment is chosen to be of the form
\begin{align}
	H_{e-p}&=\sigma_{1}^{+}\sigma_{1}^{-}\otimes\sum_{k}\hbar g_{k}(\eta(b_{1k}^{\dagger}+b_{1k})+\zeta(b_{2k}^{\dagger}+b_{2k}))\nonumber\\
	&\quad+\sigma_{2}^{+}\sigma_{2}^{-}\otimes\sum_{k}\hbar g_{k}(\eta(b_{2k}^{\dagger}+b_{2k})+\zeta(b_{1k}^{\dagger}+b_{1k})),\label{eq:H_ep}
\end{align}
where the phonon mode $b_{1k}$ (or $b_{2k}$) of frequency $\omega_k$ is coupled to both sites 1 and 2 with relative coupling strengths quantified by dimensionless scaling factors $0\le\eta\le 1$ and $\zeta\equiv\sqrt{1-\eta^2}$ (or $\zeta$ and $\eta$), respectively. The electron-phonon couplings $g_k$ are modeled by a shifted Ohmic spectral density, as will be discussed below. When $\eta=1$, leading to $\zeta=0$, the phonon environment is reduced to local phonon baths, where the phonon modes $b_{jk}$ are locally coupled to site $j$, inducing spatially uncorrelated noise. When $\eta\neq 0$ and $\zeta\neq 0$, the phonon environment is reduced to a shared phonon bath, as the phonon modes are coupled to both sites 1 and 2, leading to spatially correlated noise. A schematic representation of the local and shared phonon baths is displayed in Fig.~\ref{figureindex1}.  The degree of correlations in the noise is quantified by a correlation length $\xi$ defined by $\exp(-d/\xi) = 2\eta\zeta \in \left[0,1\right]$ with $d$ denoting the distance between sites 1 and 2.

In this work, we employ the Bloch-Redfield formalism \cite{JeskeJCP2015,JeskePRA2013} to describe the dynamics of the dimer system based on the Hamiltonian above. This formalism is well suited for describing the effect of correlated fluctuations \cite{JeskeJCP2015}, including the existence of partial correlations, as summarized in Appendix A. The phonon environment is modeled by a shifted Ohmic spectral density \cite{KreisbeckJPCL2012}
\begin{equation}
{\cal J}(\omega)=\frac{\lambda}{\pi}\left(\frac{\gamma\omega}{\gamma^2+(\omega-\Omega_s)^{2}}+\frac{\gamma\omega}{\gamma^2+(\omega+\Omega_s)^{2}}\right),
\label{eq:spec}
\end{equation}
where $\lambda$ denotes the reorganization energy, defined by $\hbar\lambda=\hbar\int_{0}^{\infty}d\omega{\cal J}(\omega)\omega^{-1}$, and $\gamma$ is the bath relaxation rate. For the FMO complex, small pure dephasing rates have been identified as a condition for the oscillations of rephasing cross-peaks \cite{KreisbeckJPCL2012}. In this work, we take the shift $\Omega_s$ of the phonon spectral density to be resonant with the exciton splitting, {\it i.e.}~$\Omega_s = \left|\epsilon_2-\epsilon_1\right|$, so that the homogeneous broadening is dominated by relaxation, rather than pure dephasing, as discussed in Appendix A. In simulations, we take a fast bath relaxation rate of $\gamma=(50\,{\rm fs})^{-1}$, corresponding to a broad spectral density, to avoid vibronic effects induced by underdamped modes, such as vibronic progressions or mixing in absorption~\cite{GelzinisJCP2015} and 2D spectra. With the Bloch-Redfield equation, we simulate 2D electronic spectra in the impulsive limit with the assumption that the transition dipoles of sites 1 and 2 are mutually orthogonal, as discussed in Appendix B.


\section{Results}

\begin{figure*}[t]
	\includegraphics[width=\textwidth]{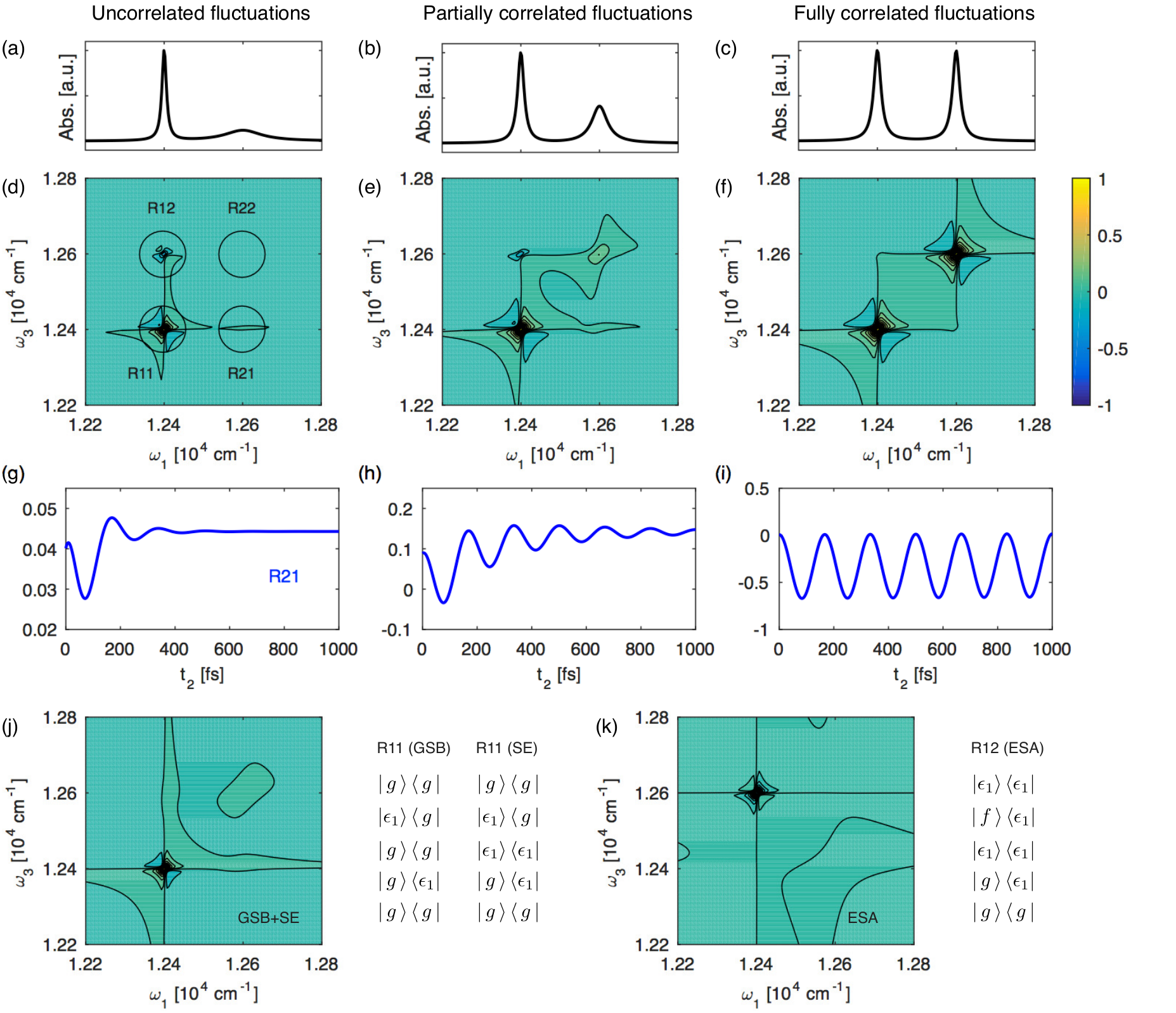}
	\caption{Absorption (Abs.), the real part of rephasing 2D spectra at waiting time $t_2=0$ and the $t_2$-transient of the lower-diagonal cross-peak R21 centered at $(\omega_1,\omega_3)=(\epsilon_2,\epsilon_1)$ with $\epsilon_1=1.24\times 10^{4}\,{\rm cm}^{-1}$ and $\epsilon_2=1.26\times 10^{4}\,{\rm cm}^{-1}$. In (a), (d), (g), (j), (k), we consider local phonon baths characterized by a short correlation length $\xi=10^{-3}d$, leading to $e^{-d/\xi}\approx 0$. In (b), (e), (h), we consider an intermediate case where $\xi=3d$, leading to $e^{-d/\xi}\approx 0.7$. In (c), (f), (i), we consider a shared phonon bath characterized by a long correlation length $\xi=10^{3}d$, leading to $e^{-d/\xi}\approx 1$. In (j), the sum of GSB and SE contributions to 2D spectra shown in (d) is displayed with the Feynman diagrams responsible for the main peak R11. In (k), the ESA contribution to (d) is displayed with the Feynman diagram for the main peak R12. The ESA contribution makes R12 stronger than R21 in both (d) and (e). Here we employed $\hbar\Omega_1=\hbar\Omega_2=12500\,{\rm cm}^{-1}$, $\hbar J=100\,{\rm cm}^{-1}$, $\hbar\lambda=50\,{\rm cm}^{-1}$, $\gamma=(50\,{\rm fs})^{-1}$ ({\it cf}.~$\hbar\gamma\approx 106\,{\rm cm}^{-1}$), $\hbar\Omega_s=200\,{\rm cm}^{-1}$ and $T=77\,{\rm K}$.}
	\label{figureindex2}
\end{figure*}

\begin{figure}[ht!]
	\includegraphics[width=0.5\textwidth]{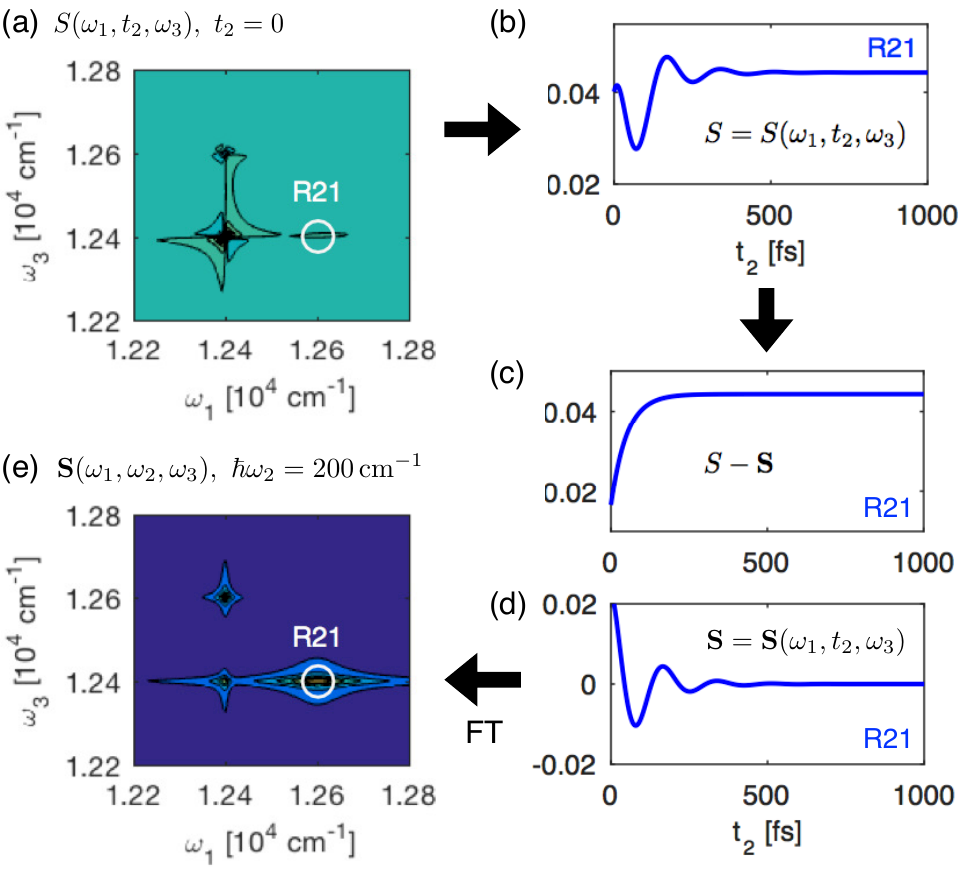}
	\caption{A schematic representation of beating map calculation. In (a), 2D spectra $S(\omega_1,t_2,\omega_3)$ at $t_2=0$ are displayed ({\it cf}.~Fig.~\ref{figureindex2}(d)). In (b), $t_2$-transient of the cross-peak R21 is shown  ({\it cf}.~Fig.~\ref{figureindex2}(g)). The transient consists of (c) non-oscillatory component $S-{\bf S}$, including exponential and static $t_2$-transients, and (d) oscillatory component ${\bf S}={\bf S}(\omega_1,t_2,\omega_3)$ (see text). By extracting the oscillatory components ${\bf S}$ from the raw 2D spectra $S$ for each $(\omega_1,\omega_3)$ value and Fourier transforming ${\bf S}$ with respect to $t_2$, one can obtain the beating map in the $(\omega_1,\omega_2,\omega_3)$ domain, as shown in (e), where the beating frequency $\omega_2$ is taken to be the exciton splitting of $\abs{\epsilon_2-\epsilon_1}$. In (a)-(d), we display the real part of $S$ for the sake of simplicity, but the imaginary part of $S$ is also included in the computation of the beating map ${\bf S}(\omega_1,\omega_2,\omega_3)$ (see text). We note that in this work, ${\bf S}^{0.1}$ is displayed ({\it cf}.~Fig.~\ref{figureindex4}(b)), instead of ${\bf S}$ ({\it cf}.~(e) and Fig.~\ref{figureindex5}(a)), to make small amplitudes more visible.}
	\label{figureindex3} 
\end{figure}

\begin{figure*}
	\includegraphics[width=\textwidth]{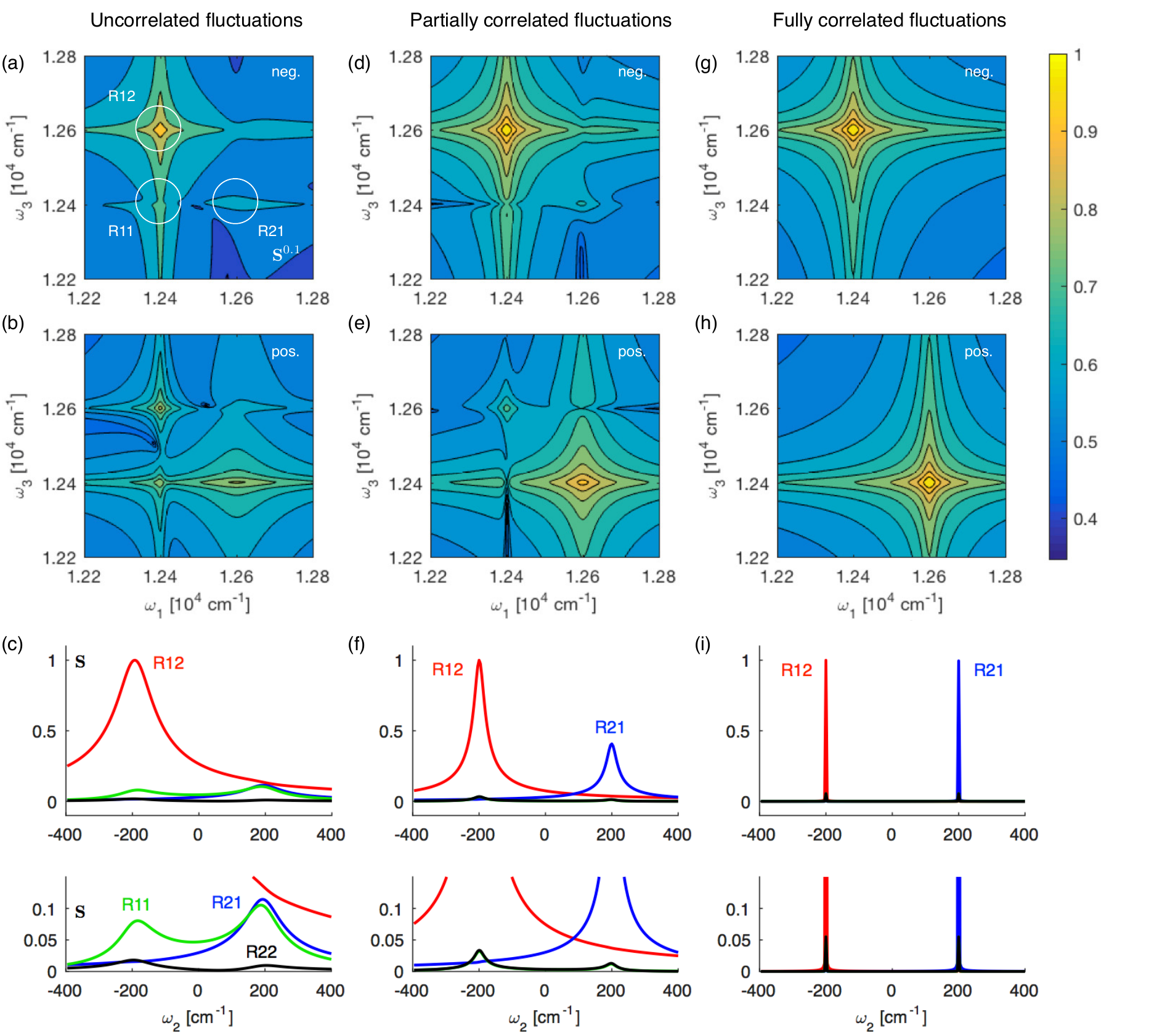}
	\caption{The beating map of complex-valued rephasing spectra that visualizes the lineshape of oscillatory 2D signals at a frequency of $\omega_2$. In (a)-(c), we consider local phonon baths with the parameters used in Fig.~\ref{figureindex2}(a). In (a), the lineshape of oscillatory signals in the form of $\exp(i\omega_2 t_2-\Gamma  t_2)$ is displayed with a negative frequency of $\hbar\omega_2=-\hbar\abs{\epsilon_2-\epsilon_1}=-200\,{\rm cm}^{-1}$ and an overall decay rate of $\Gamma $. Here the maximum value of ${\bf S}\equiv{\bf S}(\omega_1,\omega_2,\omega_3)$ is normalized to 1, {\it i.e.}~$0\le{\bf S}\le 1$, and ${\bf S}^{0.1}$ is displayed instead of ${\bf S}$, so that the small amplitudes are more visible in the beating map. In (b), the lineshape of oscillatory signals with a positive frequency of $\hbar\omega_2=\hbar\abs{\epsilon_2-\epsilon_1}=200\,{\rm cm}^{-1}$ is displayed. It is notable that the lineshape of R21 is asymmetric with a larger homogeneous broadening along $\omega_1$-axis when compared to the broadening along $\omega_3$-axis. In both (a) and (b), there are oscillations centered at the lower diagonal-peak R11. In (c), the amplitudes of the cross-peaks R12 and R21 and diagonal-peaks R11 and R22 in the beating map are displayed as a function of the beating frequency $\omega_2$. Here ${\bf S}$ is displayed instead of ${\bf S}^{0.1}$. The cross-peaks R12 and R21 are centered at negative and positive beating frequencies, respectively. On the other hand, the diagonal-peak R11 has comparable amplitudes at both positive and negative frequencies. In (d)-(f), we consider an intermediate case with the parameters used in Fig.~\ref{figureindex2}(b). In (d) and (e), the lineshape of the cross-peaks R12 and R21 is more symmetric and the diagonal-peak R11 is less visible when compared to the case of the uncorrelated noise shown in (a)-(c). In (g)-(i), we consider a shared phonon bath with the parameters used in Fig.~\ref{figureindex2}(c). In (g) and (h), the diagonal-peaks R11 and R22 are not visible, and the lineshape of the cross-peaks R12 and R21 is symmetric along the $\omega_1$- and $\omega_3$-axes. In (i), all the peaks have very narrow linewidths along $\omega_2$-axis, as the overall decay rate $\Gamma $ of the oscillatory 2D signals is very low due to the highly correlated noise ({\it cf}.~Fig.~\ref{figureindex2}(i)). In (i), the diagonal-peaks R11 and R22 have very small amplitudes, but this is due to the homogeneous broadening of the cross-peaks R12 and R21 along $\omega_1$- and $\omega_3$-axes, as the diagonal-peaks are not visible in (g) and (h).}
	\label{figureindex4} 
\end{figure*}


In this section, we provide a beating map analysis of rephasing and non-rephasing spectra of a model dimer system to demonstrate how non-secular environmental couplings and correlated fluctuations affect the oscillatory features in 2D spectra.  We will consider two cases in the simulations: a homodimer, where both sites 1 and 2 have the same site energy $\hbar\Omega_{1,2}=12500\,$cm$^{-1}$, and a heterodimer, where the two sites have different site energies, $\hbar\Omega_1 = 12600\,$cm$^{-1}$ and $\hbar\Omega_2 = 12400\,$cm$^{-1}$.  For both cases, the electronic coupling between sites is taken to be $\hbar J = 100\,$cm$^{-1}$, and the phonon bath is modeled by typical values encountered in natural photosynthetic systems \cite{IshizakiPCCP2010}: $\hbar\lambda = 50\,$cm$^{-1}$, $\gamma = (50\,$fs$)^{-1}$ and $T=77\,$K.  We will also show how the beating map is changed by static disorder.  The numerical results, which will be provided in this section, are investigated analytically in Appendix D in full detail.

Before we provide a beating map analysis, we demonstrate in Fig.~\ref{figureindex2} that correlated fluctuations can modify absorption and 2D lineshapes when homogeneous broadening is dominated by relaxation, and induce long-lived 2D oscillations within the Bloch-Redfield formalism.  For the homodimer, Fig.~\ref{figureindex2}(a) shows the absorption spectrum when correlations in the noise are negligible ({\it i.e.}~local phonon baths). The high-energy absorption peak centered at $\epsilon_2=1.26\times 10^{4}\,{\rm cm}^{-1}$ is broader than the low-energy peak centered at $\epsilon_1=1.24\times 10^{4}\,{\rm cm}^{-1}$. This is due to the fact that the homogeneous broadening is dominated by relaxation in the model, where the broadening of the low-energy peak is dominated by pure dephasing only, while the high-energy peak is broadened by both relaxation and pure dephasing. A larger broadening of the high-energy peak makes its amplitude smaller than the low-energy peak. As shown in Figs.~\ref{figureindex2}(b) and (c), the amplitude and homogeneous broadening of two peaks become similar as correlations in the noise increase, where the broadening is dominated by pure dephasing with suppressed relaxation.

In Fig.~\ref{figureindex2}(d), the real part of the rephasing spectra at waiting time $t_2=0$ is displayed for the case that correlations in the noise are absent. The high-energy diagonal peak R22 ({\it cf}.~four peaks marked by black circles in Fig.~\ref{figureindex2}(d)) is hardly visible, as expected from the small amplitude of the high-energy absorption peak shown in Fig.~\ref{figureindex2}(a). The amplitude of upper-diagonal cross-peak R12 is larger than lower-diagonal cross-peak R21. This is due to the excited state absorption (ESA) signals. In Fig.~\ref{figureindex2}(j), the sum of ground state bleaching (GSB) and stimulated emission (SE) contributions to the rephasing spectra is displayed where the main peak is R11, as its amplitude and broadening along excitation and detection axes are governed by the lineshape of the low-energy absorption peak ({\it cf.}~Feynman diagrams in Fig.~\ref{figureindex2}(j)). In Fig.~\ref{figureindex2}(k), the ESA contribution is displayed where the main peak is R12 centered at detection frequency of $\omega_3=\epsilon_2$. Here the coherence $\ket{f}\bra{\epsilon_1}$ between bi-exciton and low-energy exciton state leads to the ESA peak centered at $\omega_3=\epsilon_f-\epsilon_1=\epsilon_2$, as described in the Feynman diagram in Fig.~\ref{figureindex2}(k), leading to a large amplitude and a narrow linewidth along $\omega_3$-axis, similar to the low-energy absorption peak. Figs.~\ref{figureindex2}(e) and (f) show that the rephasing lineshape becomes more symmetric and high-energy diagonal peak R22 starts to have a larger amplitude as correlations in the noise increase.

Figs.~\ref{figureindex2}(g)-(i) show how the dynamics of lower-diagonal cross-peak R21 are affected by the degree of correlations in the noise. For uncorrelated noise, the cross-peak R21 shows oscillatory dynamics up to $t_2\approx 300\,{\rm fs}$, as shown in Fig.~\ref{figureindex2}(g). As the correlation length $\xi$ increases, the lifetime of the oscillations in R21 is increased as shown in Figs.~\ref{figureindex2}(h) and (i), describing partially and fully correlated noise, respectively. These results demonstrate that correlations in the noise can enhance the lifetime of excited state coherences, as expected from Refs.~\onlinecite{NalbachNJP2010,IshizakiNJP2010,ChenJCP2010,McCutcheonPRB2011,LimNJP2014,JeskeJCP2015,AbramaviciusJCP2011}, leading to persistent oscillatory 2D signals when $\xi/d \rightarrow \infty$. The advantage of the current formalism is that we can tune $\xi$ to cover the two extreme cases of completely uncorrelated and perfectly correlated noise.

To investigate in detail how correlations in the noise affect the oscillatory 2D signals, we use {\it a beating map analysis}, which visualizes the lineshape of oscillatory 2D signals in the $(\omega_1,\omega_3)$ domain as a function of the beating frequency $\omega_2$. To this end, we extract oscillatory components from the total 2D spectra that contain both damped oscillations and non-oscillatory components. The non-oscillatory components include exponential and static $t_2$-transients ({\it cf}.~Figs.~\ref{figureindex2}(g) and (h)). In 2D experiments, the oscillatory components are extracted from raw 2D spectra by fitting multi-exponentials to the raw $t_2$-transients for each $(\omega_1,\omega_3)$ value, or by fitting 2D decay-associated spectra (2DDAS) to the raw 2D data~\cite{RomeroNP2014,Fuller2014,MilotaJPCA2013,LimNC2015}. In this work, we directly calculate the oscillatory components by removing time-evolution operator components leading to non-oscillatory 2D signals, which will be detailed in Appendix D.  By avoiding the fitting procedure in simulations, one can avoid potential artefacts and numerical errors in the beating map.  Throughout this work, the response function that only contains the oscillatory components is denoted by ${\bf S}(\omega_1,t_2,\omega_3)$, while the total response function that contains both oscillatory and non-oscillatory signals is represented by $S(\omega_1,t_2,\omega_3)$. The oscillatory component ${\bf S}(\omega_1,t_2,\omega_3)$ is generally expressed as a sum of complex-valued damped oscillations, {\it i.e.}~${\bf S}(\omega_1,t_2,\omega_3)=\sum_{k}A_{k}(\omega_1,\omega_3) e^{(iv_k-\Gamma_k) t_2}$ with frequencies $v_k$ and associated damping rates $\Gamma_k$. We evaluate the beating map by Fourier transforming ${\bf S}(\omega_1,t_2,\omega_3)$ with respect to the waiting time $t_2$
\begin{equation}
	{\bf S}(\omega_1,\omega_2,\omega_3)=\abs{\int_{0}^{\infty}dt_2{\bf S}(\omega_1,t_2,\omega_3)\exp(-i\omega_2 t_2)},
\end{equation}
where $\omega_2$ is the beating frequency. Here we consider a complex-valued response function ${\bf S}(\omega_1,t_2,\omega_3)$, rather than only its real or imaginary part, so that we retain the full information of the oscillatory signals. In this way, we can distinguish positive and negative frequency components that oscillate in the form of $\exp(i\abs{v} t_2-\Gamma t_2)$ and $\exp(-i\abs{v} t_2-\Gamma t_2)$, respectively, with an overall decay rate of $\Gamma$. The positive and negative frequency components are reflected in the beating map as the Lorentzian functions centered at $\omega_2=\abs{v}$ and $\omega_2=-\abs{v}$, respectively, with a width of $\Gamma$ along $\omega_2$-axis. A schematic representation of the beating map evaluation is shown in Fig.~\ref{figureindex3}. A separate analysis of positive and negative frequency components has been employed to distinguish electronic and vibrational coherences for a model quantum dot system~\cite{SeibtJPCC2013} and experimentally estimate the Hamiltonian and decoherence rates of an atomic vapour~\cite{LiNC2013}.

In Fig.~\ref{figureindex4}, we show the resulting beating map of the rephasing spectra of {\it a homodimer} with the parameters used in Fig.~\ref{figureindex2}. Figs.~\ref{figureindex4}(a)-(c) show the case of local phonon baths considered in Figs.~\ref{figureindex2}(d) and (g). In Fig.~\ref{figureindex4}(a), the beating map at a negative frequency of $\omega_2=-\abs{\epsilon_2-\epsilon_1}$ is displayed, which is dominated by the upper-diagonal cross-peak R12. This is due to the inter-excitonic coherence in the form of $\ket{\epsilon_2}\bra{\epsilon_1}$, where $\ket{\epsilon_1}$ and $\ket{\epsilon_2}$ denote lower and higher energy single exciton states, respectively. The inter-excitonic coherence leads to the negative frequency component, as $\epsilon_2>\epsilon_1$. It is notable that there are weak diagonal-peaks centered at R11 and R22, which are not artefacts of the beating map calculations. Here the maximum value of ${\bf S}\equiv{\bf S}(\omega_1,\omega_2,\omega_3)$ is normalized to 1, {\it i.e.}~$0\le{\bf S}\le 1$, and ${\bf S}^{0.1}$ is displayed instead of ${\bf S}$, so that the small amplitudes are more visible in the beating map. In Fig.~\ref{figureindex4}(b), the beating map at a positive frequency of $\omega_2=\abs{\epsilon_2-\epsilon_1}$ is shown, where the lower-diagonal cross-peak R21 is induced by the inter-excitonic coherence in the form of $\ket{\epsilon_1}\bra{\epsilon_2}$. Interestingly, all the other peaks R11, R12 and R22 are visible in Fig.~\ref{figureindex4}(b) and the amplitude of R11 is comparable to that of R21. 

\begin{figure}
	\includegraphics[width=0.5\textwidth]{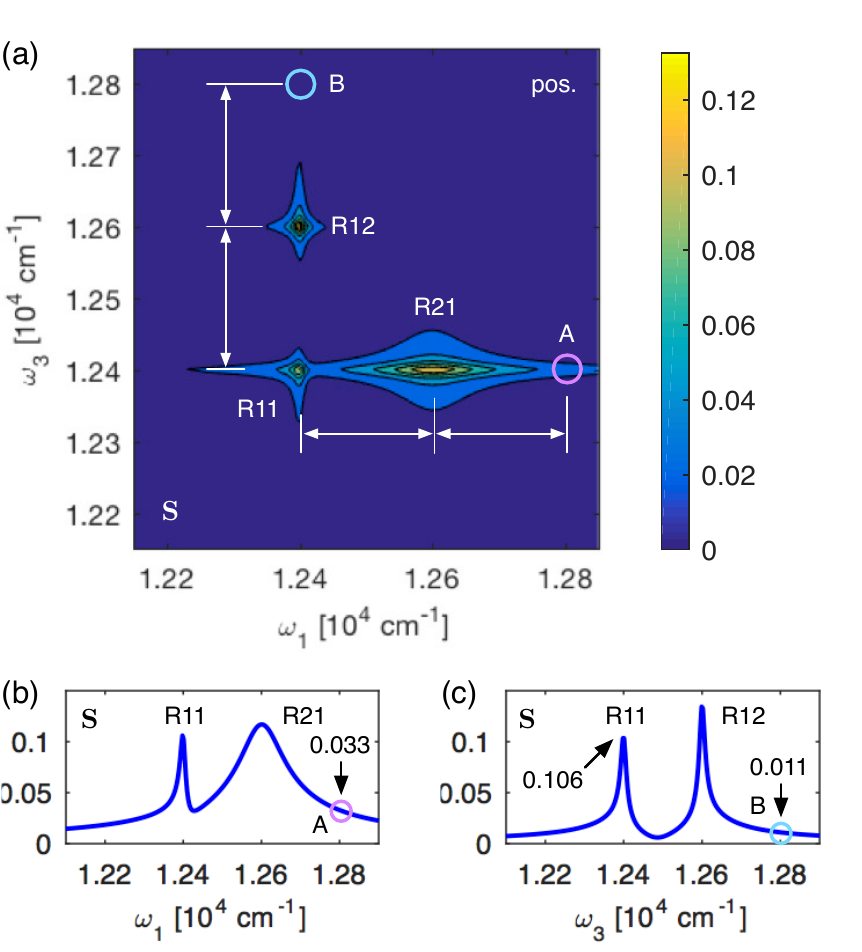}
	\caption{The cross sections of the rephasing beating map shown in Fig.~\ref{figureindex4}(b). In (a), ${\bf S}$ is displayed instead of ${\bf S}^{0.1}$ ({\it cf}.~Fig.~\ref{figureindex4}(b)). To demonstrate that the oscillations in the diagonal-peak R11 do not originate from the overlap of the homogeneous broadening of the cross-peaks R12 and R21, in (b), ${\bf S}$ is shown as a function of $\omega_1$ for $\omega_3=\epsilon_1=1.24\times 10^{4}\,{\rm cm}^{-1}$, while in (c), ${\bf S}$ is displayed as a function of $\omega_3$ for $\omega_1=\epsilon_1=1.24\times 10^{4}\,{\rm cm}^{-1}$. The sum of the values of ${\bf S}$ at points A and B is $\sim 0.044$, which is more than two times smaller than the value of R11 ($\sim 0.106$). This implies that if one assumes that the cross-peaks R12 and R21 have the Lorentzian lineshapes, the amplitude of the diagonal-peak R11 cannot be explained by the overlap of the homogeneously broadened cross-peaks R12 and R21.}
	\label{figureindex5} 
\end{figure}

\begin{figure*}
	\includegraphics[width=\textwidth]{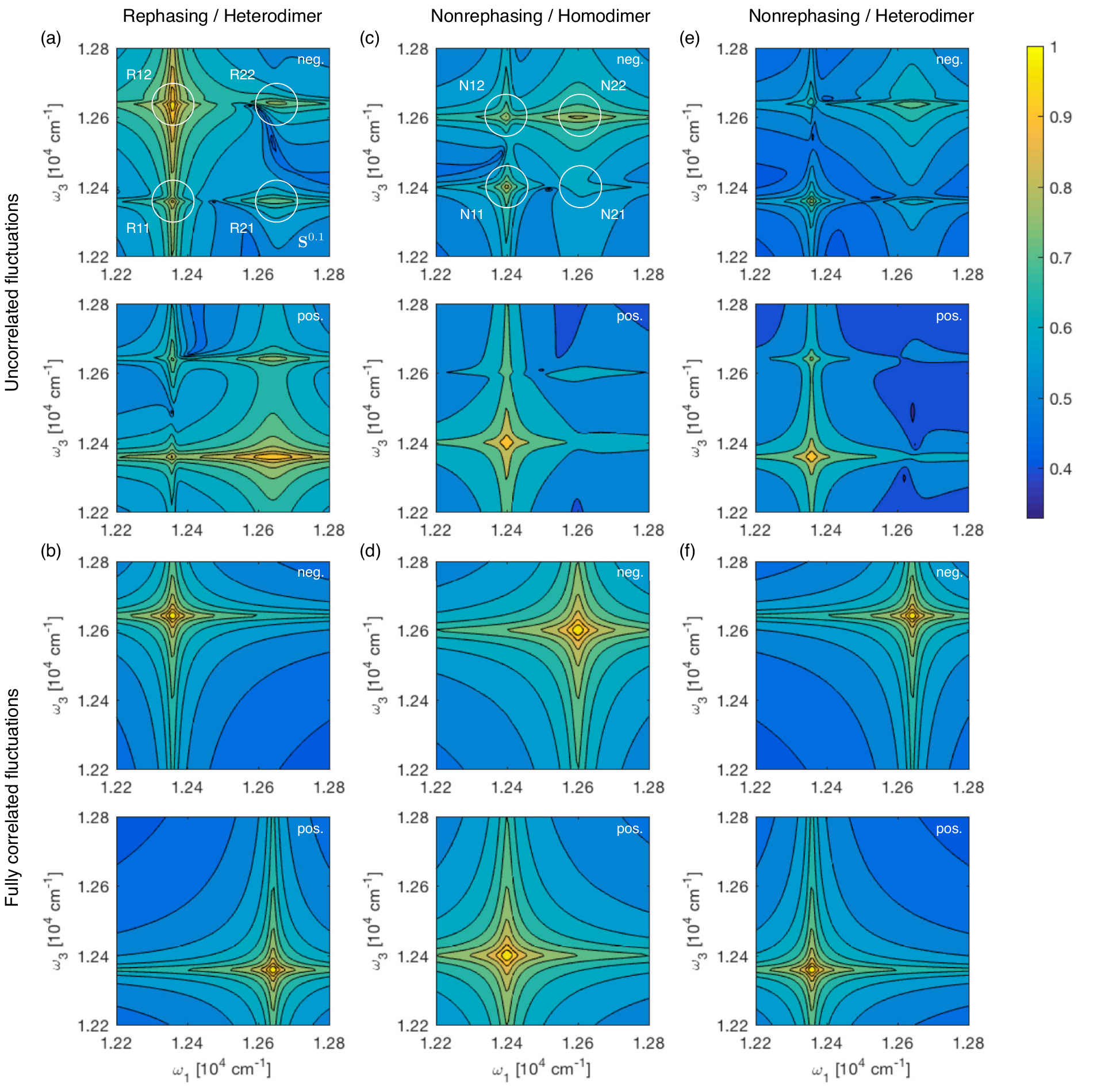}
	\caption{The rephasing beating map of a heterodimer and the non-rephasing beating map of homo- and heterodimers. In (a) and (b), we consider a heterodimer modeled by $\hbar\Omega_1=12600\,{\rm cm}^{-1}$, $\hbar\Omega_2=12400\,{\rm cm}^{-1}$, $\hbar J=100\,{\rm cm}^{-1}$, $\hbar\lambda=50\,{\rm cm}^{-1}$, $\gamma=(50\,{\rm fs})^{-1}$, $\hbar\Omega_s=\hbar\abs{\epsilon_2-\epsilon_1}\approx 283\,{\rm cm}^{-1}$ and $T=77\,{\rm K}$. In (a) and (b), the correlation length is taken to be $\xi=10^{-3}d$ (local phonon baths) and $\xi=10^{3}d$ (a shared phonon bath), respectively, for which the rephasing beating maps at negative and positive frequencies $\omega_2=\mp\abs{\epsilon_2-\epsilon_1}$ are displayed. In (c) and (d), where $\xi=10^{-3}d$ and $\xi=10^{3}d$, respectively, the non-rephasing beating maps of a homodimer are displayed with the model parameters used in Fig.~\ref{figureindex2}. In (e) and (f), where $\xi=10^{-3}d$ and $\xi=10^{3}d$, respectively, the non-rephasing beating maps of a heterodimer are displayed with the model parameters used in (a) and (b).}
	\label{figureindex6}
\end{figure*}

To understand the lineshape of oscillatory signals in more detail, Fig.~\ref{figureindex4}(c) displays the amplitudes of the cross- and diagonal-peaks as a function of the beating frequency $\omega_2$. The cross-peaks R12 and R21 are centered at the negative and positive beating frequencies, respectively. The large amplitude and broadening of the R12 peak explains the reason why R12 is visible in both Figs.~\ref{figureindex4}(a) and (b). Interestingly, the diagonal-peak R11 has comparable amplitudes at both positive and negative frequencies, contrary to the cross-peaks R12 and R21. We note that the diagonal-peak R11 in the beating map does not originate from the overlap of the homogeneously broadened cross-peaks R12 and R21. As shown in Fig.~\ref{figureindex5}(a), where ${\bf S}$ is displayed instead of ${\bf S}^{0.1}$, the homogeneous broadening of the cross-peaks is not large enough to dominate the amplitude of the diagonal-peak R11. More specifically, the distance between R11 and R21 is the same to that between R21 and point A, marked by a purple circle. Since the regime around point A has no overlap with other peaks, the amplitude of the homogeneously broadened cross-peak R21 at the position of R11 can be approximately estimated by the value of ${\bf S}$ at point A ({\it cf}.~Fig.~\ref{figureindex5}(b)). Similarly, the amplitude of the homogeneously broadened cross-peak R12 at the position of R11 can be approximately estimated by the value of ${\bf S}$ at point B, marked by a light blue circle ({\it cf}.~Fig.~\ref{figureindex5}(c)). The contribution of the cross-peaks R12 and R21 to the amplitude of R11 is more than two times smaller than the amplitude of R11, implying that the diagonal-peak does not originate solely from the overlap of the cross-peaks.

So far we have analyzed the beating map for the case that correlations in the noise are absent ({\it cf.}~Figs.~\ref{figureindex4}(a)-(c)). We now show how {\it correlations in the noise} change features of the beating map ({\it cf.}~Figs.~\ref{figureindex4}(d)-(i)). Figs.~\ref{figureindex4}(d)-(f) show the beating map in the presence of partially correlated noise ({\it cf}.~Figs.~\ref{figureindex2}(e) and (h)), while Figs.~\ref{figureindex4}(g)-(i) display the case of fully correlated noise ({\it cf}.~Figs.~\ref{figureindex2}(f) and (i)). It is notable that the overall 2D lineshapes become more symmetric as correlations in the noise increase. For instance, the asymmetric lineshape of the cross-peak R21 elongated along $\omega_1$-axis becomes more symmetric as the correlation length $\xi$ increases, as shown in Figs.~\ref{figureindex4}(b), (e) and (h). Note also that the amplitude of the diagonal peak R11 is suppressed as the correlation length $\xi$ increases.  Indeed, R11 is not visible at all for fully correlated noise, as shown in Figs.~\ref{figureindex4}(g) and (h). These results demonstrate that uncorrelated noise can induce oscillations in the rephasing diagonal-peaks and make the lineshapes of the rephasing cross-peaks asymmetric in the beating map. Conversely, correlations in the noise suppress these features, leading to symmetric lineshapes of the rephasing cross-peaks in the beating map with suppressed diagonal oscillations.

In Fig.~\ref{figureindex6}, we now show that the characteristics of the rephasing beating map of a {\it heterodimer} is similar to that of the homodimer, shown in Fig.~\ref{figureindex4}. We also show that the qualitative features of {\it non-rephasing} beating maps are significantly affected by correlations in the noise, as is the case for rephasing beating maps. Fig.~\ref{figureindex6}(a) shows rephasing beating maps of the heterodimer in the absence of correlations in the noise at negative and positive frequencies $\omega_2=\mp\abs{\epsilon_2-\epsilon_1}$. Here the oscillations occur at both cross-peaks R12 and R21 as well as at the diagonal-peaks R11 and R22. Note that the amplitude of the upper diagonal-peak R22 is more visible when compared to the case of the homodimer shown in Fig.~\ref{figureindex4}(b). Fig.~\ref{figureindex6}(b) shows the rephasing beating maps in the presence of highly correlated noise, where the oscillatory 2D signals occur only at the cross-peaks, as in the case of the homodimer. Figs.~\ref{figureindex6}(c) and (d) show the non-rephasing beating maps of the homodimer considered in Fig.~\ref{figureindex4}. In the absence of correlations in the noise, oscillatory non-rephasing signals occur at both diagonal-peaks N11 and N22 as well as at the cross-peaks N12 and N21, as shown in Fig.~\ref{figureindex6}(c). In the presence of highly correlated noise, the oscillations occur only at the diagonal-peaks N11 and N22, as shown in Fig.~\ref{figureindex6}(d). Note that the asymmetric lineshape of the diagonal-peak N22 becomes more symmetric as correlations in the noise increase. The non-rephasing beating map of a heterodimer shows similar features, as demonstrated in Figs.~\ref{figureindex6}(e) and (f). In Fig.~\ref{figureindex6}(e), the diagonal-peak N22 shows a seemingly discontinuous lineshape due to the interference of the oscillatory signals from the SE and ESA contributions, where each contribution leads to continuous 2D lineshapes in the beating map (not shown here).

These results demonstrate that in the absence of correlations in the noise, the oscillations in rephasing and non-rephasing spectra can appear at both cross- and diagonal-peaks with asymmetric lineshapes in the beating map. These features are suppressed as correlations in the noise increase, leading to oscillatory signals centered only at rephasing cross-peaks and non-rephasing diagonal-peaks with symmetric lineshapes. This is contrary to the simulated 2D spectra based on the Redfield equation within the secular approximation, where the oscillations occur only at rephasing cross-peaks and non-rephasing diagonal-peaks, even in the absence of correlations in the noise~\cite{ButkusCPL2012}. This suggests that the non-secular environmental couplings in the Bloch-Redfield equation, which couple the dynamics of an inter-excitonic coherence to that of the other inter-excitonic coherences and exciton populations~\cite{JeskeJCP2015}, may be responsible for our observations, as suggested in Refs.~\onlinecite{AbramaviciusJCP2010,PanitchayangkoonPNAS2011} for rephasing spectra without a quantitative description.  We note that our observations are not sensitive to model parameters, as shown in Appendix C and Fig.~\ref{figureindex7}, where uncorrelated static disorder is taken into account in 2D simulations. For small static disorder with a full width at half maximum (FWHM) of $50\,$cm$^{-1}$, asymmetric 2D lineshapes, rephasing diagonal-peak and non-rephasing cross-peak oscillations are visible in simulations.  For larger static disorder with a FWHM of $100\,$cm$^{-1}$, 2D lineshapes start to be elongated along the diagonal due to inhomogeneous broadening, but rephasing diagonal-peak and non-rephasing cross-peak oscillations are still visible. In Appendix D, we investigate this issue analytically to clarify how the non-secular terms and correlations in the noise affect oscillatory features in 2D spectra and how the non-secular effects can be described quantitatively with Feynman diagrams. For the homodimer, we show that non-secular oscillations are induced by non-secular interaction between different inter-excitonic coherences, as the dynamics of exciton populations are decoupled from those of coherences. For the heterodimer, where the populations and coherences are all coupled to one another, we show that non-secular oscillations are mainly induced by the mixing of different coherences, rather than population-coherence mixing, contrary to the suggestions in Refs.~\onlinecite{AbramaviciusJCP2010,PanitchayangkoonPNAS2011}.


\section{Discussion}

In Ref.~\onlinecite{LimNC2015}, three of the present authors demonstrated that the experimentally observed asymmetric lineshape in the rephasing beating map of J-aggregates cannot be explained by a correlated fluctuation model within the secular approximation. The asymmetric lineshape of the rephasing cross-peak of J-aggregates was found to originate from the fast population relaxation from higher to lower energy excitons~\cite{LimNC2015}. The present results based on the Bloch-Redfield equation beyond the secular approximation further support the claim that when the oscillatory 2D signals have a long lifetime, and the lineshapes in the beating map are sufficiently asymmetric, long-lived beating signals are not dominated by correlated fluctuations. In Ref.~\onlinecite{LimNC2015}, it was found that homogeneous broadening dominates the 2D lineshapes of J-aggregates and the exciton splitting is of the order of $\sim 700\,{\rm cm}^{-1}$. For such a large exciton splitting, non-secular effects are unlikely to induce notable signatures in oscillatory 2D signals, which are in line with the results shown in Ref.~\onlinecite{Yuen-ZhouACSN2014} based on quantum process tomography.

In Ref.~\onlinecite{CassetteNC2015}, the experimentally measured 2D spectra of colloidal semiconductor nanoplatelets were reported, where heavy- and light-hole excitons exhibit lower and higher energy peaks in 2D spectra. It was found that the 2D lineshapes of the semiconductor system are dominated by homogeneous broadening, and the broadening of the higher energy exciton is approximately three times larger than that of the lower energy exciton, leading to asymmetric 2D lineshapes in the $(\omega_1,t_2,\omega_3)$ domain: a beating map analysis in the $(\omega_1,\omega_2,\omega_3)$ domain was not provided in Ref.~\onlinecite{CassetteNC2015}. Given that highly asymmetric lineshapes were observed in experiments, our theoretical study predicts that highly correlated noise is unlikely to be present in the semiconductor system, and purely electronic coherences are unlikely to induce long-lived 2D oscillations. This is in line with the experimental observations where the lifetime of oscillatory 2D signals is similar to that of the optical coherences of the heavy- and light-hole excitons~\cite{CassetteNC2015}. The authors of Ref.~\onlinecite{CassetteNC2015} concluded that partially correlated noise is present in their system. We note that the exciton splitting of the semiconductor system was found to be in the range of $1200\sim1600\,{\rm cm}^{-1}$~\cite{CassetteNC2015}, depending on the sample preparation. For such a large exciton splitting, our results predict that non-secular effects are unlikely to induce notable features in oscillatory 2D signals. This is in line with the experimental 2D spectra where oscillatory features are present only at rephasing cross-peaks and non-rephasing diagonal-peaks~\cite{CassetteNC2015}.

In Refs.~\onlinecite{AbramaviciusJCP2010} and \onlinecite{PanitchayangkoonPNAS2011}, it was suggested that the non-secular interaction between exciton populations and inter-excitonic coherences may induce oscillatory features in the rephasing diagonal-peaks of the photosystem II reaction center and the FMO complex. Our results support the claim that electronic coherences can induce such diagonal oscillations in the rephasing spectra, mediated by non-secular couplings, as the exciton splittings of the photosynthetic systems are relatively small, typically in the range of $100\sim200\,{\rm cm}^{-1}$\cite{AbramaviciusJCP2010,PanitchayangkoonPNAS2011}. However, our quantitative analysis demonstrates that non-secular effects may be dominated by the interaction between inter-excitonic coherences, rather than the mixing of exciton populations and inter-excitonic coherences, depending on the model parameters. A detailed quantitative analysis with a beating map may be helpful for the identification of the microscopic origin of the oscillatory 2D signals of photosynthetic complexes, at least for simulated 2D spectra. We note that in Refs.~\onlinecite{AbramaviciusJCP2010} and \onlinecite{PanitchayangkoonPNAS2011}, it was suggested that non-secular terms may be related to the functional relevance of the inter-excitonic coherences in exciton transport, as these non-secular terms couple the dynamics of the inter-excitonic coherences to that of exciton populations. Our results demonstrate that non-secular effects are suppressed as correlations in the noise increase, which suggests the possibility that there could be a trade-off between non-secular effects and the lifetime of purely electronic coherences. A further theoretical investigation based on non-Markovian quantum master equations~\cite{Roden2011,Iles-Smith2016}, and numerically exact methods such as TEDOPA~\cite{PriorPRL2010,ChinNP2013} and hierarchical equations of motion (HEOM)~\cite{TanimuraJPSJ2006,StrumpferJCP2011,MascherpaarXiv2016}, could be helpful for the identification of the trade-off relation.

In our simulations, each time interval, {\it i.e.}~$t_1$, $t_2$, $t_3$, of the response function was described independently using the Bloch-Redfield equation, with the Born-Markov approximation where the phonon bath is in its equilibrium state in the electronic ground state manifold for all times. We note that the non-equilibrium dynamics of the phonon bath can induce the correlations between different time intervals and lead to much richer spectral lineshapes beyond the Lorentzians~\cite{MancalCPL12,OlsinaCP12}. In the spatially correlated noise model, slow bath relaxation or a strong coupling of electronic states to a phonon environment can induce notable temporal correlations, which can be studied using numerically exact methods, such as HEOM~\cite{KreisbeckJPCL2012,TanimuraJPSJ2006,StrumpferJCP2011}. Such a study of temporal and spatial correlations is beyond the scope of this work. Temporal correlations also play an important role in a vibronic model where underdamped vibrational modes modulate 2D lineshapes~\cite{NemethCPL08}. We also note that absorption lineshapes computed by using exact methods can show quantitative differences from those computed by using approximate methods~\cite{GelzinisJCP2015,DinhJCP2015,MaJCP2015}. Exact simulations of absorption and 2D spectra will be helpful for fully characterizing the signatures of correlated fluctuations. Such studies, which go well beyond the scope of the present work, will be pursued in a forthcoming work.


\section{Summary and Conclusions}

In this work,  we investigated the influence of non-secular couplings and spatial noise correlations on oscillatory 2D signals in rephasing and non-rephasing spectra. We employed the Bloch-Redfield formalism where we can tune the degree of correlations in the noise, such that we can cover the two extreme cases of uncorrelated and fully correlated noise.  We performed a beating map analysis to identify the signatures of non-secular effects and noise correlations in oscillatory 2D spectra.

For uncorrelated noise, we found that non-secular couplings induce the mixing of exciton populations and inter-excitonic coherences, which lead to oscillations centered at rephasing diagonal-peaks and non-rephasing cross-peaks. With a developed quantitative method, we showed that the mixing of different inter-excitonic coherences is mainly responsible for the 2D oscillations induced by non-secular couplings.  We also showed that the uncorrelated noise can induce asymmetric lineshapes of 2D peaks elongated along the excitation or detection axis.

For correlated noise, we showed that the non-secular effects are suppressed by correlations in the noise.  This spatially correlated noise can induce long-lasting 2D oscillations centered at rephasing cross- and non-rephasing diagonal-peaks, but with suppressed oscillatory features in rephasing diagonal- and non-rephasing cross-peaks.  We also showed that correlations in the noise enforce symmetry onto 2D lineshapes, hinting that the degree of asymmetry in 2D lineshapes could be used to estimate to what degree the noise is spatially correlated.  Our results demonstrate that a detailed analysis of the oscillatory features in 2D electronic spectra may provide information on the structure of vibrational environments, such as correlations in the noise and the strength of non-secular environmental couplings.


\section*{Acknowledgements}
This work was supported by the EU STREP project PAPETS and QUCHIP, the ERC Synergy grant BioQ, the Deutsche Forschungsgemeinschaft (DFG) within the SFB/TRR21 and an Alexander von Humboldt Professorship, and the state of Baden-W{\"u}rttemberg through bwHPC. This research was undertaken with the assistance of resources from the National Computational Infrastructure (NCI), which is supported by the Australian Government.


\appendix
\section{Bloch-Redfield equation}
 The Bloch-Redfield equation is expressed as
 \begin{align}
 \frac{d\rho}{dt}
 &=-\frac{i}{\hbar}[H_e,\rho]+\sum_{j,k=1}^{2}\left(-s_{j} V q_{jk} V^{\dagger} \rho\right.\label{eq:A1}\\
 &\quad\left.+V q_{jk} V^{\dagger} \rho s_{j} - \rho V \hat{q}_{jk} V^{\dagger} s_{j} + s_{j} \rho V \hat{q}_{jk} V^{\dagger}\right),\nonumber
 \end{align}
 where $\rho$ denotes the density matrix of the dimer system, $V=\sum_{n=1}^{4}\ket{\omega_n}\bra{a_n}$ is a unitary operator with $\{\ket{\omega_n}\}$ representing the electronic eigenstates of the system Hamiltonian $H_e$, defined by $H_e\ket{\omega_n}=\hbar\omega_n\ket{\omega_n}$, in an arbitrary basis $\{\ket{a_n}\}$, and the other terms are given by
 \begin{align}
s_{1}&=\sigma_{1}^{+}\sigma^-_{1},\\
s_{2}&=\sigma_{2}^{+}\sigma^-_{2},\\
\bra{a_n}q_{jk}\ket{a_m} &=\bra{a_n}V^{\dagger}s_{k} V\ket{a_m}\frac{1}{2}C_{jk}(\omega_m-\omega_n),\label{eq:A4}\\
\bra{a_n}\hat{q}_{jk}\ket{a_m} &=\bra{a_n}V^{\dagger}s_{k} V\ket{a_m}\frac{1}{2}C_{kj}(\omega_n-\omega_m),\label{eq:A5}
 \end{align}
 where the spectral functions $C_{jk}(\omega)$ are defined by
 \begin{align}
 &C_{jk}(\omega)=\frac{1}{\hbar^{2}}\int_{-\infty}^{\infty}d\tau e^{i\omega\tau}\left\langle e^{iH_{p}\tau/\hbar}B_{j}e^{-iH_{p}\tau/\hbar}B_{k} \right\rangle,\\
 &B_{1}=\sum_{k}\hbar g_{k}(\eta(b_{1k}^{\dagger}+b_{1k})+\zeta(b_{2k}^{\dagger}+b_{2k})),\\
 &B_{2}=\sum_{k}\hbar g_{k}(\eta(b_{2k}^{\dagger}+b_{2k})+\zeta(b_{1k}^{\dagger}+b_{1k})).
 \end{align}
 More specifically, we consider the site basis, given by $\ket{a_1}=\ket{g_1,g_2}$, $\ket{a_2}=\ket{e_1,g_2}$, $\ket{a_3}=\ket{g_1,e_2}$, $\ket{a_4}=\ket{e_1,e_2}$. The spectral functions $C_{jk}(\omega)$ are reduced to
 \begin{align}
 C_{11}(\omega)&=C_{22}(\omega)=C(\omega),\\
 C_{12}(\omega)&=C_{21}(\omega)=2\eta\zeta C(\omega),\\
 C(\omega)&=\sum_{k}2\pi g_{k}^{2}\left[(n(\omega_k)+1)\delta(\omega-\omega_k)\right.\\
 &\qquad\qquad\quad\left.+n(\omega_k)\delta(\omega+\omega_k)\right],\nonumber
 \end{align}
 where $n(\omega_k)=(\exp(\hbar\omega_k/k_B T)-1)^{-1}$ is the mean phonon number of a phonon mode with a frequency of $\omega_k$ at temperature $T$, while $\delta(x)$ denotes the Dirac delta function. Based on the fact that $0\le 2\eta\zeta=2\eta\sqrt{1-\eta^2}\le 1$, we introduce a correlation length $\xi$ to quantify the degree of spatial correlations in the noise, defined by $\exp(-d/\xi)=2\eta\zeta$, where $d$ denotes the spatial distance between sites 1 and 2. When $\xi\ll d$, $C_{11}(\omega)=C_{22}(\omega)=C(\omega)$ and $C_{12}(\omega)=C_{21}(\omega)\approx 0$, leading to local (or spatially uncorrelated) noise, while when $\xi\gg d$, $C_{jk}(\omega)\approx C(\omega)$ for all $j$ and $k$, leading to fully correlated noise. An intermediate case of $\xi\sim d$ leads to partially correlated noise. The correlated noise is known to enhance the lifetime of electronic coherences in the single excitation subspace~\cite{NalbachNJP2010,IshizakiNJP2010,ChenJCP2010,McCutcheonPRB2011,LimNJP2014,JeskeJCP2015,AbramaviciusJCP2011}. This is contrary to the anti-correlated noise defined by $\zeta=-\sqrt{1-\eta^2}$ in Eq.~(\ref{eq:H_ep}), which is known to suppress the lifetime of excited state coherences~\cite{IshizakiNJP2010,LimNJP2014}. In this work, we do not consider the anti-correlated noise, as we are interested in the scenario that correlations in the noise enhance the lifetime of excited state coherences, leading to long-lived oscillatory 2D signals. Therefore, the spectral functions $C_{jk}(\omega)$ in the presence of spatially correlated noise can be summarized as
 \begin{align}
 C_{11}(\omega)&=C_{22}(\omega)=C(\omega),\\
 C_{12}(\omega)&=C_{21}(\omega)=e^{-d/\xi} C(\omega).
 \end{align}
 In the continuous limit of phonon modes, leading to a phonon bath, the spectral function $C(\omega)$ is reduced to
 \begin{equation}
 C(\omega)=
 \begin{cases}
 2\pi{\cal J(\omega)}(n(\omega)+1) & \omega> 0,\\
 2\pi{\cal J(\abs{\omega})}n(\abs{\omega}) & \omega< 0,\\
 \lim_{\omega\rightarrow 0}2\pi{\cal J(\omega)}n(\omega) & \omega=0,\label{eq:dep_rxn}
 \end{cases}
 \end{equation}
 where ${\cal J(\omega)}$ is the phonon spectral density that describes the phonon mode density weighted by the electron-phonon coupling strength $g_k$ and satisfies ${\cal J}(0)=0$. In this work, $\cal{J}(\omega)$ is modeled by a shifted Ohmic spectral density described in Eq.~(\ref{eq:spec}). The shift $\Omega_s$ of the Ohmic spectral density can make $C(0) \ll C\left(\left|\epsilon_2-\epsilon_1\right|\right)$ ({\it cf.}~Eq.~(\ref{eq:dep_rxn})), for instance, when $\Omega_s\approx\left|\epsilon_2-\epsilon_1\right|$, such that the pure dephasing rates proportional to $C(0)$ are smaller than the relaxation rate between single exciton states $\ket{\epsilon_1}$ and $\ket{\epsilon_2}$.  We note that the pure dephasing and relaxation rates are not only determined by the spectral function $C(\omega)$, but also by the system parameters of the electronic Hamiltonian, described by $s_j$ and $V$ in Eqs.~(\ref{eq:A1}),~(\ref{eq:A4}) and~(\ref{eq:A5}).
 
 \section{2D electronic spectroscopy} 
In 2D experiments, three excitation pulses interact with a molecular system and the resultant third-order optical response of the system is measured as a function of the time delays between the first and second, the second and third, and the third excitation pulse and the emitted signal from the molecular system. These time delays are called coherence time $t_1$, waiting time $t_2$ and rephasing time $t_3$, respectively. The Fourier transformation of the response function with respect to $t_1$ and $t_3$ leads to 2D spectra as a function of excitation frequency $\omega_1$ and detection frequency $\omega_3$. When the pulse duration of the excitation pulses is short enough, the excitation fields can be approximately described by the Dirac delta function in the time domain, for which the third-order optical response function can be described within the rotating wave approximation in the impulsive limit. This is equivalent to the assumption that the laser spectrum is broad enough to cover the electronic states of the dimer system in the frequency domain. When the laser spectrum is not broad enough for a given system, one needs to take into account the pulse duration explicitly in 2D simulations~\cite{BrixnerJCP2004}. For the waiting times longer than the pulse duration, it was found that the finite pulse duration mainly acts as a frequency filter~\cite{KjellbergPRB2006}.

Within the rotating wave approximation in the impulsive limit, rephasing 2D spectra are formally expressed as
\begin{align}
S_{R}(\omega_1,t_2,\omega_3)&=\int_{0}^{\infty}dt_{1}\int_{0}^{\infty}dt_{3}e^{-i(\omega_1 t_1-\omega_3 t_3)}\\
&\quad\times[R_{\rm GSB}+R_{\rm SE}-R_{\rm ESA}],\nonumber
\end{align}
where $R_{\rm GSB}$, $R_{\rm SE}$ and $R_{\rm ESA}$ denote the ground state bleaching (GSB), stimulated emission (SE) and excited state absorption (ESA) contributions to the rephasing spectra, respectively:
\begin{align}
R_{\rm GSB}(t_1,t_2,t_3)&={\rm tr}[\mu^{-}u(t_3)[\mu^{+}u(t_2)[u(t_1)[\rho_{\rm eq}\mu^{-}]\mu^{+}]]],\label{eq:RGSB}\\
R_{\rm SE}(t_1,t_2,t_3)&={\rm tr}[\mu^{-}u(t_3)[u(t_2)[\mu^{+}u(t_1)[\rho_{\rm eq}\mu^{-}]]\mu^{+}]],\label{eq:RSE}\\
R_{\rm ESA}(t_1,t_2,t_3)&={\rm tr}[\mu^{-}u(t_3)[\mu^{+}u(t_2)[\mu^{+}u(t_1)[\rho_{\rm eq}\mu^{-}]]]],\label{eq:RESA}
\end{align}
with $\rho_{\rm eq}$ representing the equilibrium state in the electronic ground state manifold, $u(t)$ is a formal representation of the propagator, determined by the Bloch-Redfield equation in this work.  Here $\mu^{\pm}$ denote the transition dipole operators of the molecular system, describing the optical transition between ground and excited states by the excitation pulses
\begin{align}
\mu^{+}&=(\hat{e}\cdot\vec{d}_{1})\sigma_{1}^{+}+(\hat{e}\cdot\vec{d}_{2})\sigma_{2}^{+},\label{eq:mup}\\
\mu^{-}&=(\hat{e}\cdot\vec{d}_{1})\sigma_{1}^{-}+(\hat{e}\cdot\vec{d}_{2})\sigma_{2}^{-},\label{eq:mum}
\end{align}
where $\hat{e}$ denotes the polarization direction of the excitation pulses, which are all assumed to be parallel in this work, while $\vec{d}_k$ represents the transition dipole moment of site $k$. In 2D simulations, we take into account the rotational averaging of the dipole moments $\vec{d}_k$ with respect to the polarization direction $\hat{e}$, as we are considering 2D measurements of an ensemble of dimers. We assume that the sites 1 and 2 have mutually orthogonal transition dipoles with the same magnitude, {\it i.e.}~$\vec{d}_{1}\cdot\vec{d}_{2}=0$ and $\vec{d}_{1}\cdot\vec{d}_{1}=\vec{d}_{2}\cdot\vec{d}_{2}$. In the GSB pathway, the system is in the ground state during waiting time $t_2$, while in the SE and ESA pathways, the system is in the single excitation subspace during $t_2$. Within our model, the oscillatory 2D signals originate only from the SE and ESA contributions, as we are not considering ground state vibrational coherences induced by underdamped vibrational motions.

Similarly, non-rephasing 2D spectra can be formally expressed as
\begin{align}
S_{N}(\omega_1,t_2,\omega_3)&=\int_{0}^{\infty}dt_{1}\int_{0}^{\infty}dt_{3}e^{i(\omega_1 t_1+\omega_3 t_3)}\\
&\quad\times[N_{\rm GSB}+N_{\rm SE}-N_{\rm ESA}],\nonumber
\end{align}
where the GSB, SE and ESA contributions are expressed as
\begin{align}
N_{\rm GSB}(t_1,t_2,t_3)&={\rm tr}[\mu^{-}u(t_3)[\mu^{+}u(t_2)[\mu^{-}u(t_1)[\mu^{+}\rho_{\rm eq}]]]],\\
N_{\rm SE}(t_1,t_2,t_3)&={\rm tr}[\mu^{-}u(t_3)[u(t_2)[u(t_1)[\mu^{+}\rho_{\rm eq}]\mu^{-}]\mu^{+}]],\\
N_{\rm ESA}(t_1,t_2,t_3)&={\rm tr}[\mu^{-}u(t_3)[\mu^{+}u(t_2)[u(t_1)[\mu^{+}\rho_{\rm eq}]\mu^{-}]]].
\end{align}

\section{Inhomogeneous broadening}

\begin{figure*}[ht!]
	\includegraphics[width=\textwidth]{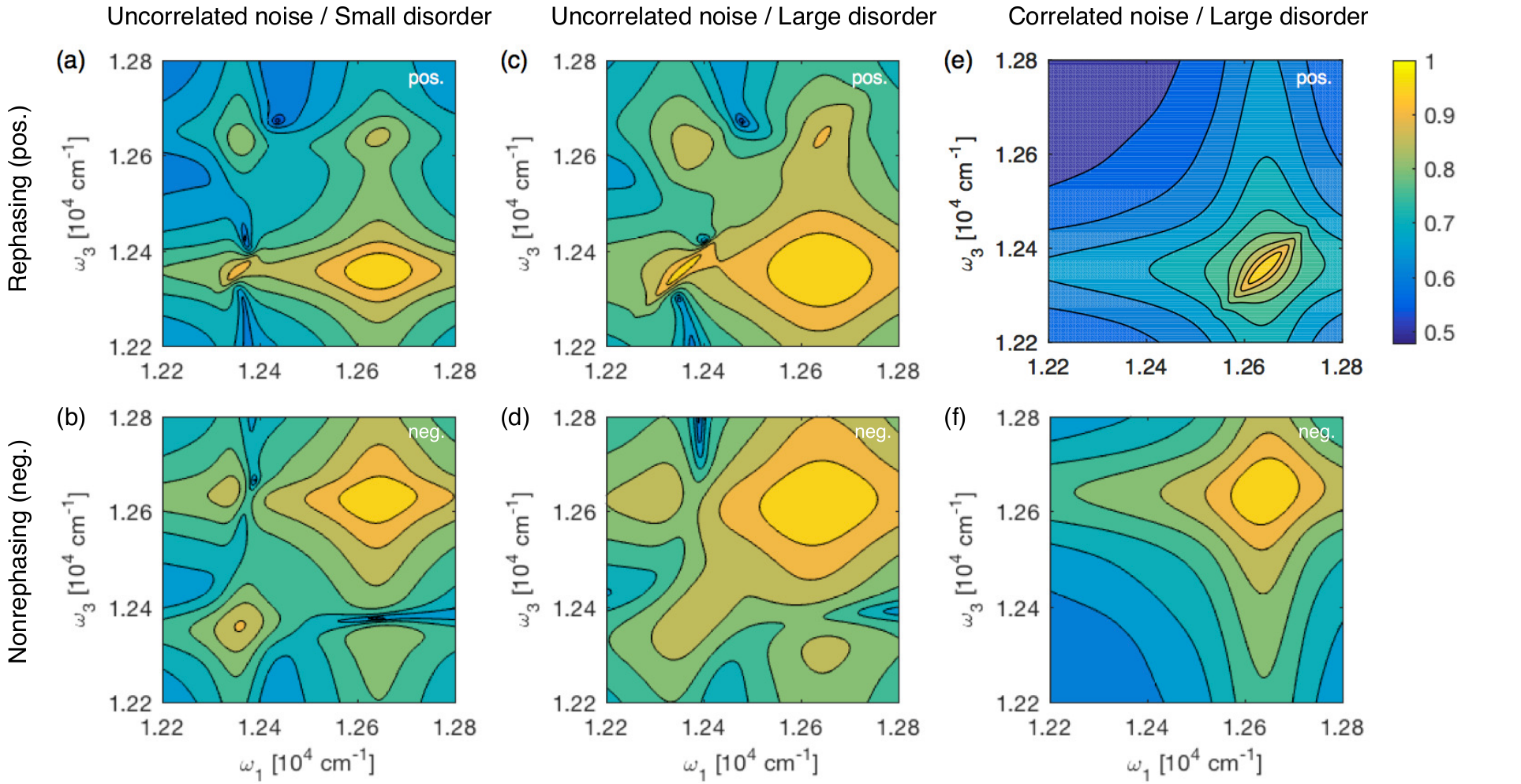}
	\caption{The rephasing and non-rephasing beating maps of a heterodimer in the presence of inhomogeneous broadening. Here we employed the model parameters used in Figs.~\ref{figureindex6}(a), (b), (e) and (f): $\hbar\mathrm{\left\langle\Omega_1\right\rangle}=12600\,{\rm cm}^{-1}$, $\hbar\mathrm{\left\langle\Omega_2\right\rangle}=12400\,{\rm cm}^{-1}$ (the average site energies), $\hbar J=100\,{\rm cm}^{-1}$, $\hbar\lambda=50\,{\rm cm}^{-1}$, $\gamma=(50\,{\rm fs})^{-1}$, $\hbar\Omega_s\approx 283\,{\rm cm}^{-1}$ (the exciton splitting for the average site energies) and $T=77\,{\rm K}$. In (a) and (b), the rephasing beating map at a positive frequency of $\hbar\omega_2=283\,{\rm cm}^{-1}$ and the non-rephasing beating map at a negative frequency of $\hbar\omega_2=-283\,{\rm cm}^{-1}$ are displayed, respectively, for the case that $\xi=10^{-3}d$ (uncorrelated fluctuations) and the inhomogeneous broadening is modeled by Gaussian distributions with a FWHM of $50\,{\rm cm}^{-1}$. In (c) and (d), the rephasing and non-rephasing beating maps are displayed, respectively, for the case that $\xi=10^{-3}d$ (uncorrelated fluctuations) and the inhomogeneous broadening is modeled by a larger FWHM of $100\,{\rm cm}^{-1}$. In (e) and (f),  the rephasing and non-rephasing beating maps are displayed, respectively, for the case that $\xi=10^{3}d$ (correlated fluctuations) and the FWHM is taken to be $100\,{\rm cm}^{-1}$.}
	\label{figureindex7} 
\end{figure*}

Here we demonstrate how oscillatory features in the beating map are affected by {\it inhomogeneous broadening}. In Fig.~\ref{figureindex7}, we consider uncorrelated disorder, where the site energies $\Omega_1$ and $\Omega_2$ of a dimer are described by two independent Gaussian distributions centered at the average values of $\mathrm{\left\langle\Omega_1\right\rangle}$ and $\mathrm{\left\langle\Omega_2\right\rangle}$, respectively. Here we consider the same full width at half maximum (FWHM) for both Gaussian distributions, and employ the model parameters of the heterodimer used in Fig.~\ref{figureindex6}, where $\hbar\mathrm{\left\langle\Omega_1\right\rangle}=12600\,{\rm cm}^{-1}$ and $\hbar\mathrm{\left\langle\Omega_2\right\rangle}=12400\,{\rm cm}^{-1}$. Figs.~\ref{figureindex7}(a) and (b) show the rephasing and non-rephasing beating maps, respectively, for the case of uncorrelated fluctuations ({\it i.e.}~$\xi=10^{-3}d$) with the inhomogeneous broadening modeled by a FWHM of $50\,{\rm cm}^{-1}$. Compared to Figs.~\ref{figureindex6}(a) and (e), where the inhomogeneous broadening is not considered, the overall 2D lineshapes in Figs.~\ref{figureindex7}(a) and (b) become broader due to the inhomogeneous broadening. However, the oscillatory features in the rephasing diagonal-peaks and non-rephasing cross-peaks ({\it i.e.}~non-secular effects) and the asymmetric 2D lineshapes elongated along $\omega_1$-axis are still visible. As the FWHM increases further, the 2D lineshapes are elongated along the diagonal ($\omega_1=\omega_3$), but non-secular effects are still visible for a FWHM of $100\,{\rm cm}^{-1}$, as shown in Figs.~\ref{figureindex7}(c) and (d).  This is somewhat relevant, as the static disorder of the FMO complex has been modeled by a FWHM of $\sim 100\,{\rm cm}^{-1}$ in other works~\cite{KreisbeckJPCL2012,AdolphsBJ2006}. On the other hand, for the case of correlated fluctuations, the rephasing beating map shows strong elongation of a cross-peak along diagonal, as shown in Fig.~\ref{figureindex7}(e), while the non-rephasing beating map shows a relatively symmetric 2D lineshape of a diagonal peak, as shown in Fig.~\ref{figureindex7}(f). This is due to the difference in the phase distributions of the rephasing and non-rephasing spectra in the $(\omega_1,\omega_3)$ domain~\cite{RancovaJCP2015,Hamm2011}.

\section{Diagonalization of the Liouville space operator}

Within the Bloch-Redfield formalism, the oscillatory signals in rephasing spectra are induced by excited state coherences described by the stimulated emission (SE) and excited state absorption (ESA) contributions in the theory of 2D spectroscopy \cite{JonasARPC2003} ({\it cf.}~Eqs.~(\ref{eq:RSE}) and (\ref{eq:RESA})). Here we show how the SE contribution to the oscillatory rephasing signals can be described quantitatively to identify the role of non-secular couplings and spatial noise correlations in the beating map. The analytical approach, which will be presented below, can be generalized to the ESA contribution to the rephasing spectra as well as to the SE and ESA contributions to the non-rephasing spectra.  The analysis is based on the diagonalization of the Liouville space operator.  This approach can be generalized to the other quantum master equations beyond the Bloch-Redfield equation employed in this work.

The lineshape of 2D spectra along excitation axis $\omega_1$ is determined by the dynamics of optical coherences between ground state and singly excited states during the coherence time $t_1$. In the exciton basis, the optical coherences are expressed as $\rho_{g1}(t_1)\ket{g}\bra{\epsilon_{1}}+\rho_{g2}(t_1)\ket{g}\bra{\epsilon_{2}}$ where the time evolution of $\rho_{g1}(t_1)$ and $\rho_{g2}(t_1)$ is governed by
\begin{equation}
\frac{d}{dt_1}
\begin{pmatrix}
	\rho_{g1}(t_1) \\ \rho_{g2}(t_1)
\end{pmatrix}
=
\begin{pmatrix}
	X_{11} & X_{12} \\ X_{21} & X_{22}
\end{pmatrix}
\begin{pmatrix}
	\rho_{g1}(t_1) \\ \rho_{g2}(t_1)
\end{pmatrix},
\label{eq:X_eq}
\end{equation}
where the super-operator $X$ describes both the Hamiltonian dynamics and decoherence.  For the Bloch-Redfield equation summarized in Appendix A, the elements of $X$ are given by
\begin{align}
	X_{11} &= -\frac{1}{4}C(0)(2-(1-e^{-d/\xi})s^{2}(2\theta))\label{eq:X11}\\
	&\quad-\frac{1}{4}C(-\Delta\epsilon)(1-e^{-d/\xi})s^{2}(2\theta)+i\epsilon_1,\nonumber\\
	X_{22} &= -\frac{1}{4}C(0)(2-(1-e^{-d/\xi})s^{2}(2\theta))\\
	&\quad-\frac{1}{4}C(\Delta\epsilon)(1-e^{-d/\xi})s^{2}(2\theta)+i\epsilon_2,\nonumber\\
	X_{12} &= \frac{1}{8}(C(0)-C(\Delta\epsilon))(1-e^{-d/\xi})s(4\theta),\label{eq:X12}\\
	X_{21} &= -\frac{1}{8}(C(0)-C(-\Delta\epsilon))(1-e^{-d/\xi})s(4\theta),\label{eq:X21}
\end{align}
with $s(\phi)\equiv\sin(\phi)$, $\theta$ quantifies the delocalization of excitons in the site basis, and $\Delta\epsilon=\abs{\epsilon_2-\epsilon_1}$ denotes the exciton splitting between $\ket{\epsilon_1}$ and $\ket{\epsilon_2}$ (see Eqs.~(\ref{eq:evec1}) and (\ref{eq:evec2})). Here $C(\omega)$ represents the spectral function determined by the phonon spectral density, as shown in Eq.~(\ref{eq:dep_rxn}).

The lineshape of 2D spectra along the excitation axis $\omega_1$ can be represented analytically by using the eigenstates of the super-operator $X$, defined by $X\vec{x}_{k}=\chi_{k}\vec{x}_{k}$. In the exciton basis, the eigenvalue equation is given by
\begin{equation}
\begin{pmatrix}
	X_{11} & X_{12} \\ X_{21} & X_{22}
\end{pmatrix}
\begin{pmatrix}
	x_{g1}^{(k)} \\ x_{g2}^{(k)}
\end{pmatrix}
=\chi_{k}
\begin{pmatrix}
	x_{g1}^{(k)} \\ x_{g2}^{(k)}
\end{pmatrix},
\end{equation}
where the eigenvector $\vec{x}_{k}$ in the Liouville space corresponds to an optical coherence $\hat{x}_{k}$ in the Hilbert space
\begin{equation}
	\hat{x}_{k}=x_{g1}^{(k)}\ket{g}\bra{\epsilon_1}+x_{g2}^{(k)}\ket{g}\bra{\epsilon_2},
\end{equation}
satisfying $\frac{d}{dt}\hat{x}_{k}=\chi_{k}\hat{x}_{k}$ with an associated eigenvalue of $\chi_{k}$. This implies that non-secular couplings $X_{12}$ and $X_{21}$ induce a mixing of two optical coherences $\ket{g}\bra{\epsilon_1}$ and $\ket{g}\bra{\epsilon_2}$ in the exciton basis. The dynamics of the mixed coherence is formally described by $u(t_1)[\hat{x}_k]=e^{\chi_k t_1}\hat{x}_k$.  The optical coherence created by the first excitation pulse, {\it i.e.} $\ket{g}\bra{g}\mu^-$, can be represented as a superposition of $\hat{x}_k$
\begin{equation}
	\ket{g}\bra{g}\mu^{-}=\sum_{j=1}^{2}\ket{g}\mathrm{\langle}\epsilon_j\mathrm{|}\mu_{gj}=\sum_{k=1}^{2}\alpha_{k}\hat{x}_{k},
	\label{eq:1st_coh}
\end{equation}
where $\mu_{gj}$ represents the transition dipole strength between ground state $\ket{g}$ and the $j$-th exciton $\mathrm{|}\epsilon_j\mathrm{\rangle}$ for a given realization of the transition dipole moments of sites 1 and 2 ({\it cf}.~Eqs.~(\ref{eq:mup}) and (\ref{eq:mum})). The coefficient $\alpha_k$ in Eq.~(\ref{eq:1st_coh}) describes the effective transition dipole strength between ground state $\ket{g}\bra{g}$ and mixed coherence $\hat{x}_{k}$, given by
\begin{equation}
\begin{pmatrix}
	\alpha_{1} \\
	\alpha_{2}
\end{pmatrix}
=
\begin{pmatrix}
	x^{(1)}_{g1} & x^{(2)}_{g1} \\
	x^{(1)}_{g2} & x^{(2)}_{g2}
\end{pmatrix}^{-1}
\begin{pmatrix}
	\mu_{g1}\\
	\mu_{g2}
\end{pmatrix}.
\end{equation}
The dynamics of $\ket{g}\bra{g}\mu^{-}$ during time $t_1$ is then expressed as
\begin{equation}
	u(t_1)[\ket{g}\bra{g}\mu^{-}]=\sum_{k=1}^{2}\alpha_{k}e^{\chi_{k}t_1}\hat{x}_{k},
	\label{eq:gecoh}
\end{equation}
and the Fourier transformation of $e^{\chi_{k}t_1}$ determines the lineshape of the homogeneously broadened 2D spectra along the excitation axis $\omega_1$
\begin{equation}
	\int_{0}^{\infty}dt_1 e^{-i\omega_1 t_1}u(t_1)[\ket{g}\bra{g}\mu^{-}]=\sum_{k=1}^{2}\frac{\alpha_{k}}{\chi_{k}-i\omega_1}\hat{x}_{k}.
\end{equation}
Here the real and imaginary parts of the eigenvalue $\chi_{k}$ of $\hat{x}_k$, denoted by ${\rm Re}[\chi_{k}]$ and ${\rm Im}[\chi_{k}]$, respectively, determine the homogeneous broadening and peak location, respectively, of the $k$-th Lorentzian peak along the excitation axis. The non-secular couplings $X_{12}$ and $X_{21}$ in Eq.~(\ref{eq:X_eq}) can make the imaginary part of $\chi_k$ deviate from the eigenvalue $\epsilon_k$ of the system Hamiltonian ({\it cf.}~Eqs.~(\ref{eq:e1}) and (\ref{eq:e2})), implying that 2D peak locations can be shifted by non-secular effects.

These results imply that the dynamics of the eigenstates $\hat{x}_1$ and $\hat{x}_2$ of the super-operator $X$ lead to the lower and higher energy peaks, respectively, along the excitation axis ({\it cf.}~Figs.~\ref{figureindex2}-\ref{figureindex6}). When the off-diagonal components $X_{12}$ and $X_{21}$ are comparable or larger in magnitude than the difference in the diagonal components $X_{11}$ and $X_{22}$, {\it e.g.}~$\abs{X_{12}}\gtrsim\abs{X_{11}-X_{22}}$, the eigenstates $\hat{x}_k$ become a superposition of the optical coherences $\ket{g}\bra{\epsilon_1}$ and $\ket{g}\bra{\epsilon_2}$ in the exciton basis. We note that the difference in $X_{11}$ and $X_{22}$ is larger in magnitude than the exciton splitting, {\it i.e.}~$\abs{X_{11}-X_{22}}\ge\abs{{\rm Im}[X_{11}-X_{22}]}=\abs{\epsilon_2-\epsilon_1}$, indicating that, as expected, the non-secular effects are suppressed as the exciton splitting increases. The mixing is also suppressed by correlations in the noise, {\it i.e.}~$X_{11}$, $X_{22} \neq 0$ and $X_{12}$, $X_{21}\rightarrow0$ as $\left(1-e^{-d/\xi}\right)\rightarrow 0$ in Eqs.~(\ref{eq:X11})-(\ref{eq:X21}).

So far we have analyzed the dynamics of the optical coherences created by the first excitation pulse. We now consider the populations and coherences within the single excitation subspace created by the second excitation pulse.  The population or coherence in the excited state manifold is expressed in the exciton basis as
\begin{equation}
	\sum_{i,j=1}^{2}\rho_{ij}(t_2)\ket{\epsilon_i}\mathrm{\langle}\epsilon_j\mathrm{|},
\end{equation}
whose dynamics are governed by the super-operator $Y$,
\begin{equation}
\frac{d}{dt_2}
\begin{pmatrix}
	\rho_{11}(t_2) \\ \rho_{22}(t_2) \\ \rho_{12}(t_2) \\ \rho_{21}(t_2)
\end{pmatrix}
=
\begin{pmatrix}
	Y_{11,11} & Y_{11,22} & Y_{11,12} & Y_{11,21} \\
	Y_{22,11} & Y_{22,22} & Y_{22,12} & Y_{22,21} \\
	Y_{12,11} & Y_{12,22} & Y_{12,12} & Y_{12,21} \\
	Y_{21,11} & Y_{21,22} & Y_{21,12} & Y_{21,21}
\end{pmatrix}
\begin{pmatrix}
	\rho_{11}(t_2) \\ \rho_{22}(t_2) \\ \rho_{12}(t_2) \\ \rho_{21}(t_2)
\end{pmatrix}.
\end{equation}
For the Bloch-Redfield equation, the elements $Y_{jk,lm}$ of $Y$ are given by
\begin{widetext}
\begin{align}
\begin{pmatrix}
	Y_{11,11} & Y_{11,22} & Y_{11,12} & Y_{11,21} \\
	Y_{22,11} & Y_{22,22} & Y_{22,12} & Y_{22,21} \\
	Y_{12,11} & Y_{12,22} & Y_{12,12} & Y_{12,21} \\
	Y_{21,11} & Y_{21,22} & Y_{21,12} & Y_{21,21}
\end{pmatrix}
&=
\begin{pmatrix}
	0 & 0 & 0 & 0 \\
	0 & 0 & 0 & 0 \\
	0 & 0 & -i(\epsilon_1-\epsilon_2) & 0 \\
	0 & 0 & 0 & -i(\epsilon_2-\epsilon_1) \\
\end{pmatrix}
+
\frac{1-e^{-d/\xi}}{4}
\left(
\begin{array}{cc}
	-2C(-\Delta\epsilon)s^{2}(2\theta) & 2C(\Delta\epsilon)s^{2}(2\theta) \\
	2C(-\Delta\epsilon)s^{2}(2\theta) & -2C(\Delta\epsilon)s^{2}(2\theta) \\
	C(-\Delta\epsilon)s(4\theta) & -C(\Delta\epsilon)s(4\theta) \\
	C(-\Delta\epsilon)s(4\theta) & -C(\Delta\epsilon)s(4\theta) \\
\end{array}
\right.\nonumber\\
&
\left.
\begin{array}{cc}
	C(0)s(4\theta) & C(0)s(4\theta) \\
	-C(0)s(4\theta) & -C(0)s(4\theta) \\
	-(C(\Delta\epsilon)+C(-\Delta\epsilon))s^{2}(2\theta)-4C(0)c^{2}(2\theta) & (C(\Delta\epsilon)+C(-\Delta\epsilon))s^{2}(2\theta) \\
	(C(\Delta\epsilon)+C(-\Delta\epsilon))s^{2}(2\theta) & -(C(\Delta\epsilon)+C(-\Delta\epsilon))s^{2}(2\theta)-4C(0)c^{2}(2\theta)
\end{array}
\right),\label{eq:Y}
\end{align}
\end{widetext}
with $c(\phi)\equiv\cos(\phi)$. The first term on the right hand side in Eq.~(\ref{eq:Y}) shows the Hamiltonian contribution to the system dynamics, which is proportional to the exciton splitting of $\epsilon_2-\epsilon_1$. This implies that as the exciton splitting increases in magnitude, the difference in diagonal components of $Y$ increases, and as a result the non-secular interactions between exciton populations $\rho_{ii}(t)$ and inter-excitonic coherences $\rho_{ij}(t)$ with $i\neq j$, and those between different inter-excitonic coherences $\rho_{12}(t)$ and $\rho_{21}(t)$ are suppressed. The second term on the right hand side in Eq.~(\ref{eq:Y}) describes decoherence within the single excitation subspace. The factor $(1-e^{-d/\xi})$ in Eq.~(\ref{eq:Y}) shows that in the long correlation length limit, {\it i.e.}~$\xi\gg d$, there is no decoherence in the single excitation subspace, and the dynamics of the singly excited states are governed by the Hamiltonian only. The dynamics of these singly excited states can be described by the eigenstates $\hat{y}_l$ of the super-operator $Y$, satisfying 
\begin{equation}
\begin{pmatrix}
	Y_{11,11} & Y_{11,22} & Y_{11,12} & Y_{11,21} \\
	Y_{22,11} & Y_{22,22} & Y_{22,12} & Y_{22,21} \\
	Y_{12,11} & Y_{12,22} & Y_{12,12} & Y_{12,21} \\
	Y_{21,11} & Y_{21,22} & Y_{21,12} & Y_{21,21}
\end{pmatrix}
\begin{pmatrix}
	y_{11}^{(l)} \\ y_{22}^{(l)} \\ y_{12}^{(l)} \\ y_{21}^{(l)}
\end{pmatrix}
=
\upsilon_l
\begin{pmatrix}
	y_{11}^{(l)} \\ y_{22}^{(l)} \\ y_{12}^{(l)} \\ y_{21}^{(l)}
\end{pmatrix},
\end{equation}
which is expressed in the exciton basis as
\begin{equation}
	\hat{y}_l=\sum_{i,j=1}^{2}y_{ij}^{(l)}\ket{\epsilon_i}\mathrm{\langle}\epsilon_j\mathrm{|}.
\end{equation}
The time evolution of the eigenstate $\hat{y}_l$ is formally expressed as $u(t_2)[\hat{y}_l]=e^{\upsilon_l t_2}\hat{y}_l$ where the real and imaginary parts of the eigenvalue $\upsilon_l$ describe the decay and phase evolution, respectively, of the eigenstate $\hat{y}_l$. The phase evolution leads to oscillatory 2D signals. For the dimer system considered in simulations, we found that two of the eigenvalues $\upsilon_l$ have imaginary parts, which are approximately given by ${\rm Im}[\upsilon_1]\approx\abs{\epsilon_2-\epsilon_1}$ and ${\rm Im}[\upsilon_2]\approx-\abs{\epsilon_1-\epsilon_2}$. The associated two eigenstates $\hat{y}_1$ and $\hat{y}_2$ are responsible for the oscillatory 2D signals with positive and negative frequencies, respectively. The other eigenstates $\hat{y}_3$ and $\hat{y}_4$ have negligible imaginary parts, implying that they are responsible for non-oscillatory 2D signals, such as exponential and static $t_2$-transients. The time evolution of the SE contribution during the waiting time $t_2$ can be expressed as
\begin{align}
	u(t_2)[\mu^{+}u(t_1)[\ket{g}\bra{g}\mu^{-}]]&=\sum_{k=1}^{2}\alpha_{k}e^{\chi_{k}t_1}u(t_2)[\mu^{+}\hat{x}_{k}]\\
	&=\sum_{k=1}^{2}\alpha_{k}e^{\chi_{k}t_1}\sum_{l=1}^{4}\beta_{kl}e^{\upsilon_l t_2}\hat{y}_{l},\label{eq:Y_evolution}
\end{align}
where $\mu^{+}\hat{x}_{k}=\sum_{l=1}^{4}\beta_{kl}\hat{y}_{l}$ with $\beta_{kl}$ representing the effective transition dipole strength between eigenstates $\hat{x}_{k}$ and $\hat{y}_{l}$. In 2D simulations, one can calculate the beating map directly by removing the non-oscillatory components $\hat{y}_3$ and $\hat{y}_4$ from Eq.~(\ref{eq:Y_evolution}), then Fourier transforming $e^{\upsilon_l t_2}$ which leads to the $l$-th Lorentzian peak along the $\omega_2$-axis ({\it cf.}~$l=1,2$).

Finally we consider the dynamics of the optical coherences $\rho_{1g}(t_3)\ket{\epsilon_1}\bra{g}+\rho_{2g}(t_3)\ket{\epsilon_2}\bra{g}$ created by the third excitation pulse, whose dynamics during rephasing time $t_3$ are described by
\begin{equation}
\frac{d}{dt_3}
\begin{pmatrix}
	\rho_{1g}(t_3) \\ \rho_{2g}(t_3)
\end{pmatrix}
=
\begin{pmatrix}
	X_{11}^{*} & X_{12}^{*} \\ X_{21}^{*} & X_{22}^{*}
\end{pmatrix}
\begin{pmatrix}
	\rho_{1g}(t_3) \\ \rho_{2g}(t_3)
\end{pmatrix}.
\end{equation}
The eigenstates of the super-operator $X^{*}$ are given by
\begin{equation}
	\hat{x}_{k}^{*}=\left(x_{g1}^{(k)}\right)^{*}\ket{\epsilon_1}\bra{g}+\left(x_{g2}^{(k)}\right)^{*}\ket{\epsilon_2}\bra{g},
\end{equation}
with the associated eigenvalues $\chi_{k}^{*}$.  Here $\chi_{k}^{*}$ (or $\hat{x}_{k}^{*}$) is the complex conjugate (or adjoint) of $\chi_k$ (or $\hat{x}_k$). The SE contribution to the rephasing spectra (cf.~Eq.~(\ref{eq:RSE})) is then expressed as
\begin{align}
	R_{\rm SE}(t_1,t_2,t_3)=\sum_{k,m=1}^{2}\sum_{l=1}^{4}\alpha_{k}e^{\chi_{k}t_1}\beta_{kl}e^{\upsilon_l t_2}\gamma_{lm}e^{\chi_{m}^{*}t_3}{\rm tr}[\mu^{-}\hat{x}_{m}^{*}],
\end{align}
where $\hat{y}_{l}\mu^{+}=\sum_{m=1}^{2}\gamma_{lm}\hat{x}_{m}^{*}$ and $\gamma_{lm}$ denotes the effective transition dipole strength between eigenstates $\hat{y}_{l}$ and $\hat{x}_{m}^{*}$. The summations over $k,l,m$, where $k\in\{1,2\}$, $l\in\{1,2,3,4\}$ and $m\in\{1,2\}$, lead to 16 different Feynman pathways for the SE contribution to the rephasing spectra. Since only $\hat{y}_{1}$ and $\hat{y}_{2}$ are responsible for oscillatory 2D signals, there are only eight Feynman pathways with $l\in\{1,2\}$ contributing to the beating map. Thus, the SE contribution to the oscillatory rephasing signals in the $(\omega_1,\omega_3)$ domain can be expressed as
\begin{equation}
	{\bf R}_{\rm SE}(\omega_1,t_2,\omega_3)=\sum_{k,l,m=1}^{2}A_{klm}(\omega_1,\omega_3)e^{\upsilon_l t_2},
\end{equation}
where the two-dimensional amplitude $A_{klm}(\omega_1,\omega_3)$ describes a Lorentzian peak centered at $(\omega_1,\omega_3)=({\rm Im}[\chi_k],-{\rm Im}[\chi_{m}^{*}])$, weighted by the effective transition dipole strength
\begin{equation}
	A_{klm}(\omega_1,\omega_3)=\frac{\left\langle\alpha_{k}\beta_{kl}\gamma_{lm}{\rm tr}[\mu^{-}\hat{x}_{m}^{*}]\right\rangle}{(\chi_k-i\omega_1)(\chi_{m}^{*}+i\omega_3)},
\end{equation}
where the homogeneous broadenings along $\omega_1$- and $\omega_3$-axes are determined by the real part of the eigenvalues $\chi_k$ and $\chi_{m}^{*}$, respectively.  Here $\left\langle\alpha_{k}\beta_{kl}\gamma_{lm}{\rm tr}[\mu^{-}\hat{x}_{m}^{*}]\right\rangle$ denotes the rotational average (ensemble) of the effective transition dipole strength ({\it cf.}~Appendix B). These results show that $\hat{y}_{l=1,2}$ can induce oscillatory rephasing signals centered at $(\omega_1,\omega_3)=({\rm Im}[\chi_{k}],-{\rm Im}[\chi_{m}^{*}])$ when the associated transition dipole strength $\left<\alpha_k \beta_{kl} \gamma_{lm}\right.$tr$\left.\left[\mu^- \hat{x}_m^{*}\right] \right>$ is not zero.

When the optical coherences $\hat{x}_{k=1,2}$ are approximately represented by $\hat{x}_{k=1,2}\approx \ket{g}\bra{\epsilon_k}$, the eigenstates $\hat{y}_{l=1,2}$ can induce 2D oscillations centered at rephasing lower diagonal-peak R11 when the Feynman pathway of $\ket{g}\bra{g}\rightarrow\hat{x}_1\rightarrow\hat{y}_l\rightarrow\hat{x}_1^{*}\rightarrow\ket{g}\bra{g}$ has a non-zero transition dipole strength.  Here the transition from $\hat{x}_1 \approx \ket{g}\bra{\epsilon_1}$ to $\hat{y}_l$ is allowed when $\hat{y}_l\ket{\epsilon_1}\neq0$, and the transition from $\hat{y}_l$ to $\hat{x}_1^{*}\approx\ket{\epsilon_1}\bra{g}$ is allowed when $\bra{\epsilon_1}\hat{y}_l\neq0$.  These conditions are not satisfied within the secular approximation \cite{JeskeJCP2015} where $\hat{y}_1=\ket{\epsilon_1}\bra{\epsilon_2}$ and $\hat{y}_2 = \ket{\epsilon_2}\bra{\epsilon_1}$ and the super-operator $Y$ is approximated by 
\begin{equation}
\frac{d}{dt_2}
\begin{pmatrix}
\rho_{11}(t_2) \\ \rho_{22}(t_2) \\ \rho_{12}(t_2) \\ \rho_{21}(t_2)
\end{pmatrix}
=
\begin{pmatrix}
Y_{11,11} & Y_{11,22} & 0 & 0 \\
Y_{22,11} & Y_{22,22} & 0 & 0 \\
0 & 0 & Y_{12,12} & 0 \\
0 & 0 & 0 & Y_{21,21}
\end{pmatrix}
\begin{pmatrix}
\rho_{11}(t_2) \\ \rho_{22}(t_2) \\ \rho_{12}(t_2) \\ \rho_{21}(t_2)
\end{pmatrix}.
\end{equation}
On the other hand, in the presence of non-secular couplings, the eigenstates $\hat{y}_{l=1,2}$ become a mixture of different inter-excitonic coherences $\ket{\epsilon_1}\bra{\epsilon_2}$ and $\ket{\epsilon_2}\bra{\epsilon_1}$ and exciton populations $\ket{\epsilon_1}\bra{\epsilon_1}$ and $\ket{\epsilon_2}\bra{\epsilon_2}$, which are formally expressed as $\hat{y}_{l}=y^{(l)}_{12}\ket{\epsilon_1}\bra{\epsilon_2}+y^{(l)}_{21}\ket{\epsilon_2}\bra{\epsilon_1}+y^{(l)}_{11}\ket{\epsilon_1}\bra{\epsilon_1}+y^{(l)}_{22}\ket{\epsilon_2}\bra{\epsilon_2}$.  As shown in Appendix E, the conditions of $\hat{y}_l\ket{\epsilon_1}\neq0$ and $\bra{\epsilon_1}\hat{y}_l\neq0$ can be satisfied for the homo- and heterodimers considered in our simulations. For the homodimer, non-secular interaction between populations and coherences is absent, as shown in Eq.~(\ref{eq:nonabs}), where the eigenstates $\hat{y}_{l=1,2}$ are the mixtures of different inter-excitonic coherences only, {\it i.e.} $\hat{y}_{l=1,2}=y^{(l)}_{12}\ket{\epsilon_1}\bra{\epsilon_2}+y^{(l)}_{21}\ket{\epsilon_2}\bra{\epsilon_1}$, as shown in Eqs.~(\ref{eq:y1_homo}) and (\ref{eq:y2_homo}).  In this case, the dynamics of exciton populations are decoupled from those of inter-excitonic coherences, but the non-secular interaction between coherences can induce rephasing diagonal-peak oscillations. For the heterodimer, all the exciton populations and inter-excitonic coherences are coupled to one another and induce a population-coherence mixing in the eigenstates $\hat{y}_{l=1,2}$, which also lead to oscillations centered at rephasing diagonal peaks.  We found that for the model parameters used in our simulations, the mixing is dominated by inter-excitonic coherences, and the contribution of exciton populations to $\hat{y}_{l=1,2}$ is relatively small, as shown in Eqs.~(\ref{eq:yhetero1}) and (\ref{eq:yhetero2}).  A detailed quantitative description of non-secular effects with associated Feynman diagrams is provided in Appendix E.

Finally, we note that the asymmetric lineshape in the beating map originates from the fact that the lower and higher energy peaks have different homogeneous broadenings.  When the spectral function satisfies $C\left(\Delta\epsilon\right)>C\left(0\right)$, as is the case of the model parameters used in this work, relaxation dominates the homogeneous broadening, and the higher energy peak shows a larger broadening than the lower energy peak, described by $\mathrm{|}{\rm Re}[\chi_1]\mathrm{|}<\mathrm{|}{\rm Re}[\chi_2]\mathrm{|}$.  This leads to asymmetric lineshapes in the beating map, as shown in Figs.~\ref{figureindex4}(a) and (b).  As correlations in the noise increase, the super-operator $X$ is governed by the pure dephasing noise described by $C\left(0\right)$, as shown in Eqs.~(\ref{eq:X11})-(\ref{eq:X21}), where the relaxation described by $C\left(\pm\Delta\epsilon\right)$ does not contribute to the homogeneous broadening, leading to $\mathrm{|}{\rm Re}[\chi_1]\mathrm{|}\approx\mathrm{|}{\rm Re}[\chi_2]\mathrm{|}$ and symmetric lineshapes in the beating map, as shown in Figs.~\ref{figureindex4}(g) and (h).

\section{Non-secular effects}
Here we apply the quantitative method developed in Appendix D to the model parameters of homo- and heterodimers considered in our simulations. We show that the mixing of inter-excitonic coherences is mainly responsible for the oscillations centered at rephasing diagonal peaks.

For the model parameters of the homodimer with $\xi=10^{-3}d$ ({\it i.e.}~local phonon baths), the off-diagonal terms $X_{12}$ and $X_{21}$ in Eqs.~(\ref{eq:X12}) and (\ref{eq:X21}) are zero, as $\sin(4\theta)=0$ with $\theta=\pi/4$ ({\it cf}.~Eqs.~(\ref{eq:evec1}) and (\ref{eq:evec2})). This implies that the eigenstates of the super-operator $X$ (or $X^{*}$) are given by $\hat{x}_1=\ket{g}\bra{\epsilon_1}$ and $\hat{x}_2=\ket{g}\bra{\epsilon_2}$ (or $\hat{x}_{1}^{*}=\ket{\epsilon_1}\bra{g}$ and $\hat{x}_{2}^{*}=\ket{\epsilon_2}\bra{g}$). In this case, the super-operator $Y$ is reduced to
\begin{equation}
\frac{d}{dt_2}
\begin{pmatrix}
\rho_{11}(t_2) \\ \rho_{22}(t_2) \\ \rho_{12}(t_2) \\ \rho_{21}(t_2)
\end{pmatrix}
=
\begin{pmatrix}
Y_{11,11} & Y_{11,22} & 0 & 0 \\
Y_{22,11} & Y_{22,22} & 0 & 0 \\
0 & 0 & Y_{12,12} & Y_{12,21} \\
0 & 0 & Y_{21,12} & Y_{21,21}
\end{pmatrix}
\begin{pmatrix}
\rho_{11}(t_2) \\ \rho_{22}(t_2) \\ \rho_{12}(t_2) \\ \rho_{21}(t_2)
\end{pmatrix},
\label{eq:nonabs}
\end{equation}
as $\sin(4\theta)=0$ ({\it cf}.~Eq.~(\ref{eq:Y})), which shows that the dynamics of exciton populations $\rho_{ii}(t_2)$ are decoupled from those of inter-excitonic coherences $\rho_{ij}(t_2)$ with $i\neq j$. However, the non-zero off-diagonal components $Y_{12,21}$ and $Y_{21,12}$ induce non-secular interactions between different inter-excitonic coherences $\rho_{12}(t_2)$ and $\rho_{21}(t_2)$. Even though the off-diagonal components $\hbar Y_{12,21}=\hbar Y_{21,12}\approx 53\,{\rm cm}^{-1}$ are an order of magnitude smaller than the difference in diagonal components, $\hbar\abs{Y_{12,12}-Y_{21,21}}\approx 400\,{\rm cm}^{-1}$, the eigenstates $\hat{y}_{l=1,2}$ of the super-operator $Y$ show a notable mixing of different inter-excitonic coherences:
\begin{align}
\hat{y}_1&\approx 0.991\ket{\epsilon_1}\bra{\epsilon_2}-0.133\,i \ket{\epsilon_2}\bra{\epsilon_1},\label{eq:y1_homo}\\
\hat{y}_2&\approx 0.133\,i\ket{\epsilon_1}\bra{\epsilon_2}+0.991\ket{\epsilon_2}\bra{\epsilon_1},\label{eq:y2_homo}\\
\hat{y}_3&\approx 0.717\ket{\epsilon_1}\bra{\epsilon_1} - 0.717\ket{\epsilon_2}\bra{\epsilon_2},\\
\hat{y}_4&\approx 0.999\ket{\epsilon_1}\bra{\epsilon_1} + 0.024\ket{\epsilon_2}\bra{\epsilon_2},
\end{align}
with the associated eigenvalues given by $\hbar\upsilon_1=(-53+193\,i)\,{\rm cm}^{-1}$, $\hbar\upsilon_2=(-53-193\,i)\,{\rm cm}^{-1}$, $\hbar\upsilon_3=-105\,{\rm cm}^{-1}$ and $\upsilon_4=0$. The first eigenstate $\hat{y}_1$ is a superposition of the inter-excitonic coherences $\ket{\epsilon_1}\bra{\epsilon_2}$ and $\ket{\epsilon_2}\bra{\epsilon_1}$ due to the non-secular couplings $Y_{12,21}$ and $Y_{21,12}$. The imaginary part of the associated eigenvalue ${\rm Im}[\upsilon_1]\approx 193\,{\rm cm}^{-1}$ shows that $\hat{y}_1$ leads to a positive frequency component in the beating map with a beating frequency of $\omega_2\approx 193\,{\rm cm}^{-1}$. Due to the non-secular effects, the beating frequency is slightly different from the exciton splitting of $\epsilon_2-\epsilon_1=200\,{\rm cm}^{-1}$. Similarly, the second eigenstate $\hat{y}_2$ is a superposition of the inter-excitonic coherences, but it has a larger amplitude in $\ket{\epsilon_2}\bra{\epsilon_1}$ than in $\ket{\epsilon_1}\bra{\epsilon_2}$, contrary to $\hat{y}_1$. This results in the imaginary part of the associated eigenvalue having the opposite sign, ${\rm Im}[\upsilon_2]\approx -193\,{\rm cm}^{-1}$, implying that $\hat{y}_2$ leads to a negative frequency component in the beating map with $\omega_2\approx -193\,{\rm cm}^{-1}$. The eigenvalues of the other eigenstates $\hat{y}_{3}$ and $\hat{y}_{4}$ do not contain imaginary parts, implying that they are associated with non-oscillatory 2D signals: $\upsilon_3<0$ and $\upsilon_4=0$ indicate that $\hat{y}_{3}$ describes the relaxation of exciton populations, while $\hat{y}_{4}$ is an equilibrium state within the excited state manifold.

\begin{figure}[ht!]
	\includegraphics{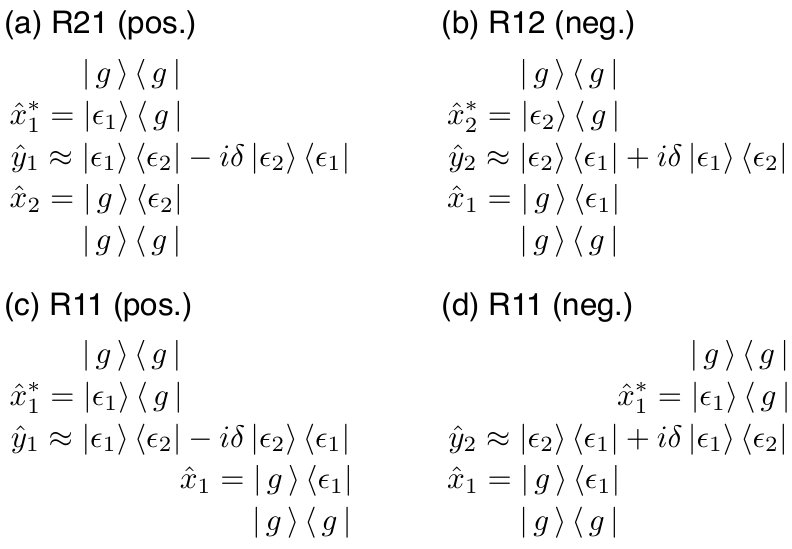}
	\caption{The Feynman diagrams of the SE contribution to the oscillatory signals in the rephasing spectra of a homodimer. In (a) and (b), the Feynman diagrams responsible for the oscillatory signals in the rephasing cross-peaks R21 and R12 are displayed, respectively. In (c) and (d), the Feynman diagrams responsible for the oscillatory signals in the rephasing diagonal peak R11 are shown, which lead to positive and negative frequency components, respectively. Here a small amplitude $\delta$, satisfying $0<\delta<1$, is employed to approximately represent the eigenstates $\hat{y}_{1}$ and $\hat{y}_{2}$ (see Eqs.~(\ref{eq:y1_homo}) and (\ref{eq:y2_homo})).}
	\label{figureindex8} 
\end{figure}

Fig.~\ref{figureindex8} shows the Feynman diagrams of the SE contribution to the oscillatory rephasing signals for the homodimer. Figs.~\ref{figureindex8}(a) and (b) show the Feynman diagrams responsible for the oscillations in the rephasing cross-peaks R21 and R12, respectively. Here the eigenstates $\hat{y}_{1}\approx\ket{\epsilon_1}\bra{\epsilon_2}-i\delta\ket{\epsilon_2}\bra{\epsilon_1}$ and $\hat{y}_{2}\approx\ket{\epsilon_2}\bra{\epsilon_1}+i\delta\ket{\epsilon_1}\bra{\epsilon_2}$ are approximately represented in terms of a small amplitude $0<\delta<1$. Note that the optical transitions $\hat{x}_{2}\rightarrow\hat{y}_{1}\rightarrow\hat{x}_{1}^{*}$ in Fig.~\ref{figureindex8}(a) and $\hat{x}_{1}\rightarrow\hat{y}_{2}\rightarrow\hat{x}_{2}^{*}$ in Fig.~\ref{figureindex8}(b) are allowed even in the absence of the small amplitude $\delta$. On the other hand, Fig.~\ref{figureindex8}(c) shows the Feynman diagram responsible for the positive frequency component in the rephasing diagonal-peak R11, where the optical transition $\hat{x}_{1}\rightarrow\hat{y}_{1}$ is allowed only if the small amplitude $\delta$ is non-zero, {\it i.e.}~$\hat{x}_{1}=\ket{g}\bra{\epsilon_1}\rightarrow-i\delta\ket{\epsilon_2}\bra{\epsilon_1}+\ket{\epsilon_1}\bra{\epsilon_2}\approx\hat{y}_{1}$, as a direct transition from $\ket{g}\bra{\epsilon_1}$ to $\ket{\epsilon_1}\bra{\epsilon_2}$ is forbidden. This implies that the mixing $\delta$ of different inter-excitonic coherences induced by the non-secular couplings $Y_{12,21}$ and $Y_{21,12}$ allows the transition $\hat{x}_{1}\rightarrow\hat{y}_{1}\rightarrow\hat{x}_{1}^{*}$ in Fig.~\ref{figureindex8}(c) to occur, leading to oscillations centered at the rephasing diagonal-peak R11. Similarly, Fig.~\ref{figureindex8}(d) shows the Feynman diagram that induces the negative frequency component in the rephasing diagonal-peak R11.

\begin{figure}[ht!]
	\includegraphics{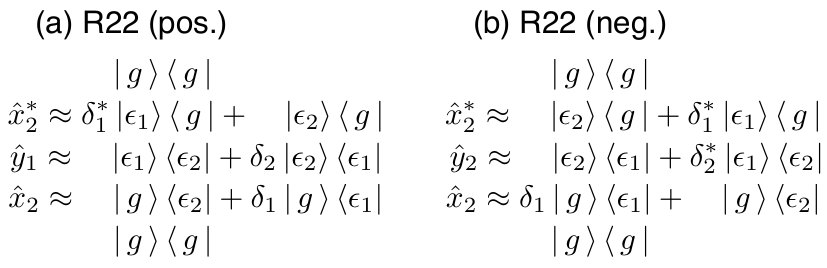}
	\caption{The Feynman diagrams of the SE contribution to the oscillatory signals in the rephasing spectra of a heterodimer. In (a) and (b), the Feynman diagrams responsible for the oscillations in the rephasing diagonal-peak R22 are displayed, which induce positive and negative frequency components, respectively. Here $\abs{\delta_1}<1$ and $\abs{\delta_2}<1$ are small amplitudes induced by non-secular effects.}
	\label{figureindex9} 
\end{figure}

We now consider the model parameters of the heterodimer with $\xi=10^{-3}d$ ({\it i.e.}~local phonon baths). In this case, the off-diagonal components $X_{12}$ and $X_{21}$ are non-zero, which makes the eigenstates $\hat{x}_{1}$ and $\hat{x}_{2}$ of the super-operator $X$ be in a superposition of the optical coherences $\ket{g}\bra{\epsilon_1}$ and $\ket{g}\bra{\epsilon_2}$ in the exciton basis, given by
\begin{align}
\hat{x}_{1}&\approx 0.999\ket{g}\bra{\epsilon_1}+0.005 e^{1.69\,i}\ket{g}\bra{\epsilon_2},\\
\hat{x}_{2}&\approx 0.993\ket{g}\bra{\epsilon_2}+0.115 e^{-1.45\,i}\ket{g}\bra{\epsilon_1}.
\end{align}
The super-operator $Y$ is reduced to
\begin{equation}
\frac{d}{dt_2}
\begin{pmatrix}
\rho_{11}(t_2) \\ \rho_{22}(t_2) \\ \rho_{12}(t_2) \\ \rho_{21}(t_2)
\end{pmatrix}
=
\begin{pmatrix}
Y_{11,11} & Y_{11,22} & Y_{11,12} & Y_{11,21} \\
Y_{22,11} & Y_{22,22} & Y_{22,12} & Y_{22,21} \\
Y_{12,11} & Y_{12,22} & Y_{12,12} & Y_{12,21} \\
Y_{21,11} & Y_{21,22} & Y_{21,12} & Y_{21,21}
\end{pmatrix}
\begin{pmatrix}
\rho_{11}(t_2) \\ \rho_{22}(t_2) \\ \rho_{12}(t_2) \\ \rho_{21}(t_2)
\end{pmatrix},
\end{equation}
which includes additional non-zero off-diagonal elements, such as $Y_{12,22}$, inducing the non-secular coupling between exciton populations and inter-excitonic coherences. The eigenstates of the super-operator $Y$ are given by
\begin{align}
\hat{y}_{1}
&\approx 0.997\ket{\epsilon_1}\bra{\epsilon_2}+0.061 e^{-1.59\,i}\ket{\epsilon_2}\bra{\epsilon_1}\label{eq:yhetero1}\\
&\quad+0.011 e^{-1.53\,i}\ket{\epsilon_1}\bra{\epsilon_1}+0.011 e^{1.61\,i}\ket{\epsilon_2}\bra{\epsilon_2},\nonumber\\
\hat{y}_{2}
&\approx 0.061 e^{1.59\,i}\ket{\epsilon_1}\bra{\epsilon_2}+0.997\ket{\epsilon_2}\bra{\epsilon_1}\label{eq:yhetero2}\\
&\quad+0.011 e^{1.53\,i}\ket{\epsilon_1}\bra{\epsilon_1}+0.011 e^{-1.61\,i}\ket{\epsilon_2}\bra{\epsilon_2},\nonumber\\
\hat{y}_{3}
&\approx 0.169 e^{1.55\,i}\ket{\epsilon_1}\bra{\epsilon_2}+0.169 e^{-1.55\,i}\ket{\epsilon_2}\bra{\epsilon_1}\\
&\quad+0.686\ket{\epsilon_1}\bra{\epsilon_1}-0.686\ket{\epsilon_2}\bra{\epsilon_2},\nonumber\\
\hat{y}_{4}
&\approx 0.999\ket{\epsilon_1}\bra{\epsilon_1}+0.005\ket{\epsilon_2}\bra{\epsilon_2},
\end{align}
with the associated eigenvalues given by $\hbar\upsilon_1\approx (-41+280\,i)\,{\rm cm}^{-1}$, $\hbar\upsilon_2\approx (-41-280\,i)\,{\rm cm}^{-1}$, $\hbar\upsilon_3\approx -70\,{\rm cm}^{-1}$ and $\upsilon_4=0$. As is the case of the homodimer, the first two eigenstates $\hat{y}_1$ and $\hat{y}_2$ are responsible for oscillatory 2D signals, while the other eigenstates $\hat{y}_3$ and $\hat{y}_4$ induce exponential and static $t_2$-transients, respectively. Note that the eigenstates are mixtures of exciton populations and inter-excitonic coherences due to non-secular effects. Fig.~\ref{figureindex9} shows the Feynman diagrams of the SE contribution to the oscillations in the rephasing diagonal-peak R22 of the heterodimer. Contrary to the case of the homodimer, the mixing of the optical coherences $\ket{g}\bra{\epsilon_1}$ and $\ket{g}\bra{\epsilon_2}$ during coherence and rephasing times, described by a small amplitude $\abs{\delta_1}<1$, enhances the diagonal oscillations in the rephasing spectra.


\end{document}